\documentclass{aa}
\usepackage[dvipsnames]{xcolor}
\usepackage{graphicx}
\usepackage{txfonts}
\usepackage{upgreek}
\usepackage{gensymb}
\usepackage{tikz}
\usepackage{courier}
\usepackage[bf,normalsize]{subfigure}
\usepackage{lscape}
\usepackage{afterpage}
\usepackage{placeins}           
\usepackage{abraces}
\usepackage[separate-uncertainty=true,table-align-uncertainty=true]{siunitx}
\usepackage{newtxtext}
\usepackage[frenchmath,varg]{newtxmath}
\usepackage{natbib,twoopt}
\usepackage[breaklinks=true,colorlinks=True,linkcolor=NavyBlue!50!Emerald,citecolor=Gray!70!ProcessBlue,filecolor=NavyBlue!50!Emerald,urlcolor=NavyBlue!50!Emerald]{hyperref}
\usepackage[para,online,flushleft]{threeparttable}
\usepackage{orcidlink}
\modulolinenumbers[10]

\bibpunct{(}{)}{;}{a}{}{,} 

\newcommand{\kms}{\,\hbox{\hbox{km}\,\hbox{s}$^{-1}$}}
\newcommand{\ergs}{\,\hbox{\hbox{erg}\,\hbox{s}$^{-1}$}}

\def\micron{\hbox{$\mu$m}}

\begin{document}

   \titlerunning{Wild is the Wind}
   \authorrunning{Fernández-Ontiveros et al.}
   \title{Wild is the wind from low-luminosity AGN: A jet-driven gas bubble blowing out a massive CO-dark outflow in ESO\,420-G13}

   \author{J.\,A.~Fernández-Ontiveros\orcidlink{0000-0001-9490-899X} \inst{\ref{CEFCA},\ref{CEFCA-UA}}\thanks{\fontsize{8.9}{9.5}\selectfont \sffamily \email{\sffamily \href{mailto:j.a.fernandez.ontiveros@gmail.com}{j.a.fernandez.ontiveros@gmail.com},\,\href{mailto:jafernandez@cefca.es}{jafernandez@cefca.es}}} \and L.~Spinoglio\orcidlink{0000-0001-8840-1551}\inst{\ref{IAPS}} \and M.~Pereira-Santaella\orcidlink{0000-0002-4005-9619}\inst{\ref{IFF}} \and A.~Hernán-Caballero\orcidlink{0000-0002-4237-5500}\inst{\ref{CEFCA},\ref{CEFCA-UA}} \and E.~Hatziminaoglou\orcidlink{0000-0003-0917-9636}\inst{\ref{IAC},\ref{ULL},\ref{ESO}} \and E.~Pérez-Montero\orcidlink{0000-0003-3985-4882}\inst{\ref{IAA}} \and J.\,M.~Vílchez\orcidlink{0000-0001-7299-8373}\inst{\ref{IAA}} \and B.~Pérez-Díaz\orcidlink{0000-0002-0939-9156}\inst{\ref{OAR}} \and R.~Amorín\orcidlink{0000-0001-5758-1000}\inst{\ref{IAA}} \and \\ M.\,A.~Malkan\orcidlink{0000-0001-6919-1237}\inst{\ref{UCLA}} \and K.\,M.~Dasyra\orcidlink{0000-0002-1482-2203}\inst{\ref{NKUA}}}

   \institute{
   Centro de Estudios de F\'isica del Cosmos de Arag\'on (CEFCA), Plaza San Juan 1, 44001 Teruel, Spain \label{CEFCA}
   \and
   Unidad Asociada CEFCA--IAA, CEFCA, Unidad Asociada al CSIC por el IAA y el IFCA, Plaza San Juan 1, 44001 Teruel, Spain \label{CEFCA-UA}
   \and
   Istituto di Astrofisica e Planetologia Spaziali (INAF--IAPS), Via Fosso del Cavaliere 100, 00133 Roma, Italy \label{IAPS}
   \and
   Instituto de Física Fundamental, CSIC, Calle Serrano 123, 28006 Madrid, Spain \label{IFF}
   \and
   Instituto de Astrofísica de Canarias (IAC), 38205 La Laguna, Tenerife, Spain \label{IAC}
   \and
   Universidad de La Laguna, Dpto. Astrofísica, 38206 La Laguna, Tenerife, Spain \label{ULL}
   \and
   European Southern Observatory, Karl-Schwarzschild-Stra{\ss}e 2, 85748, Garching, Germany \label{ESO}
   \and
   Instituto de Astrofísica de Andalucía (IAA--CSIC), Glorieta de la Astronomía s/n, 18008 Granada, Spain \label{IAA}
   \and
   INAF--Osservatorio Astronomico di Roma, via Frascati 33, 00078, Monteporzio Catone, Italy \label{OAR}
   \and
   Astronomy Division, University of California, Los Angeles, CA 90095-1547, USA \label{UCLA}
   \and
   Section of Astrophysics, Astronomy \& Mechanics, Department of Physics, National and Kapodistrian University of Athens, Panepistimioupolis Zografou, 15784 Athens, Greece \label{NKUA}
   }
   \date{Received 13 March 2026 / Accepted 9 July 2026}

\abstract{
We present JWST/MIRI mid-infrared integral field spectroscopy combined with ALMA CO(2--1) observations of the post-starburst galaxy ESO\,420-G13, hosting a low-luminosity active galactic nucleus (AGN). The unprecedented spatial and spectral resolution of MIRI enables a detailed study of the molecular and ionised gas kinematics, excitation, and energetics in the nuclear kiloparsec, revealing the impact of AGN feedback in a system with modest radiative output. Despite its faint radio and X-ray emission ($L_\mathrm{2-10keV} \sim 10^{40}\, \mathrm{erg\,s^{-1}}$), \mbox{ESO\,420-G13} exhibits powerful kinetic feedback in the form of massive molecular and ionised gas outflows, with a total kinetic power of $\sim 1.4 \times 10^{41}\, \mathrm{erg\,s^{-1}}$. This corresponds to a jet–interstellar medium (ISM) coupling efficiency of $\sim 3.8\%$, within the range observed in more powerful AGNs. The feedback is driven by a previously undetected compact jet, traced by collimated coronal-line and extended X-ray emission to $\gtrsim 870\, \mathrm{pc}$ from the nucleus. The interaction is strongest $\sim 370\, \mathrm{pc}$ north of the nucleus, where a fast ionised gas stream emerges perpendicular to the jet axis, coinciding with a bend in the jet direction. Enhanced velocity dispersion in warm H$_2$ surrounds this gas stream, consistent with an expanding molecular bubble. Massive molecular outflows are detected at its edges; the blueshifted outflow is devoid of CO emission, likely due to CO destruction in shocks or by cosmic rays from the jet--ISM interaction. About 5\% of the central molecular reservoir has already been expelled, and the remaining gas is turbulent and warm, suggesting an ongoing phase of AGN-driven feedback in this post-starburst galaxy. Our results highlight the enormous potential of mid-infrared imaging spectroscopy to uncover jet-driven feedback in low-luminosity AGNs. Without the spatially resolved MIRI diagnostics, the kinetic power of the AGN in ESO\,420-G13 and its role in shaping the host galaxy ISM would have remained hidden.}


\maketitle
\nolinenumbers

\section{Introduction}\label{intro}

Powerful feedback from active galactic nuclei (AGNs), through massive outflows and jets, is required by current simulations to reconcile the dark matter halo and galaxy stellar mass functions \citep{silk12,wechsler18,behroozi19}, explain the migration of star-forming galaxies from the blue cloud to the red sequence \citep{schawinski14,heckman14}, and reproduce several empirical relations in galaxies. These include correlations between supermassive black hole (SMBH) mass and bulge stellar velocity dispersion \citep{magorrian98,ferrarese00,kormendy13}, between stellar mass and metallicity \citep{lequeux79,tremonti04,andrews13}, the Tully-Fisher relation in spiral galaxies \citep{tully77,mcgaugh00}, and the Faber-Jackson relation in ellipticals \citep{faber76,bernardi03}.

AGN feedback is commonly divided into a `radio' or `kinetic' mode and a `quasar' or `radiative' mode (see \citealt{fabian12} for a review). The kinetic mode is powered by winds and relativistic jets that inject energy into the interstellar medium (ISM), whereas the radiative mode dominates at high Eddington ratios, when the SMBH releases most of its energy as radiation \citep{silk98,fabian99}. This radiation can heat and expel the ISM, suppressing star formation \citep{dimatteo05,somerville15}. Cosmological simulations suggest that jet feedback is the most efficient quenching mechanism, while radiatively driven winds mainly suppress star formation in intermediate-mass galaxies without strongly affecting their long-term stellar mass growth \citep{scharre24}.

Although luminous AGNs are too rare to dominate galaxy evolution, low-luminosity AGNs (LLAGNs) may represent an important and underestimated source of feedback. They are the most numerous AGNs in the nearby Universe \citep{ho08}, so their cumulative impact may exceed that of more luminous nuclei. LLAGN are characterised by low accretion rates ($L_\text{bol} / L_\text{Edd} \lesssim 10^{-3}$) and radiatively inefficient accretion \citep{narayan94}, while jet activity is expected to dominate \citep[e.g.][]{fender04}. Indeed, most faint AGNs host compact jets, usually detected as unresolved radio cores \citep{nagar05,baldi23} and occasionally accompanied by kiloparsec-scale lobes. These jets can drive massive molecular outflows \citep{aalto12,dasyra15,jafo20,audibert19} and dissipate energy through highly ionised winds, turbulence, and shocks over hundreds of parsecs \citep{holt08,rodriguez17,venturi21,goold24}. This can strongly alter the dynamics and thermal state of the warm molecular gas \citep{ps22,ogle24,lopez25}, suppressing the cooling required for star formation \citep{feruglio20}.

In this context, post-starburst galaxies are of particular interest for understanding quenching. Characterised by strong Balmer absorption lines ($\mathrm{H\delta}$ equivalent widths $\gtrsim 3$--$5$\,\AA; \citealt{poggianti09,werle22}) and weak or absent nebular emission, they are considered transition systems between star-forming and quiescent galaxies, often following starburst or merger events \citep{zabludoff96,goto05}. Although rare in the nearby Universe ($\lesssim 1$\%), they become significantly more common towards cosmic noon \citep[$\sim 5$\%;][]{wild16}. Nuclear activity is observed in a substantial fraction \citep{alatalo15,french21}, suggesting that AGN feedback may play a critical role in halting star formation, although a direct causal link is complicated by the short duty cycle of radiatively efficient AGNs relative to the longer-lived post-starburst features. Hydrodynamical simulations suggest that black hole-driven winds or jets are required to reproduce the suppressed star formation and central gas depletion observed in these systems \citep{zheng20}, with kinetic feedback expected to dominate at low accretion rates where radiative efficiency is reduced \citep[e.g.][]{alatalo11}. Post-starburst galaxies thus bridge the active and passive phases of galaxy evolution, providing critical constraints on the timescales and efficiency of AGN feedback during quenching.

The aim of this paper is to investigate AGN feedback in ESO\,420-G13 ($D = 49.4$\,Mpc\footnote{Flat $\Lambda$CDM cosmology with $H_0 = 73$\,km\,s$^{-1}\, \mathrm{Mpc^{-1}}$, $\Omega_\mathrm{m} = 0.27$, and $z = 0.01205$ \citep{jafo20}.}; $\log(M^*/\mathrm{M_\odot}) \sim 10.7$, \citealt{bi20}), a galaxy with strong post-starburst A-star signatures \citep{thomas17} hosting an X-ray-faint AGN \citep[$L_\mathrm{2-10\,keV} \sim 10^{40}\, \mathrm{erg\,s^{-1}}$;][]{lehmer10} that has launched a massive molecular outflow ($8.3 \times 10^6\, \mathrm{M_\odot}$; \citealt{jafo20}), indicating strong jet--ISM interaction. The AGN coexists with ongoing starburst activity and a largely undisturbed cold molecular gas reservoir ($3 \times 10^8\, \mathrm{M_\odot}$) in the innermost few kiloparsecs. Using \textit{JWST}/MIRI mid-IR spectroscopy and ALMA CO(2--1) observations, we characterise how AGN feedback operates in ESO\,420-G13 by tracing the propagation of mechanical energy across the multiphase ISM: from the highly ionised, shock-heated gas along the jet trail (coronal lines [\ion{Ne}{v}]$_{14.3,24.3}$, [\ion{Ne}{vi}]$_{7.7}$, [\ion{O}{iv}]$_{25.9}$), through the warm molecular phase traced by H$_2$\,S(1)--S(8) transitions, to the cold molecular disc. By combining the spatially resolved morphology and kinematics of these tracers, we quantify how and where AGN energy couples to the ISM on sub-kiloparsec scales.

The paper is organised as follows. Section\,\ref{data} presents observations and data reduction. Section\,\ref{results} describes the results, focusing on warm molecular gas (Section\,\ref{res_H2}) and highly ionised gas emission (Sections\,\ref{res_coronal} and \ref{res_stream}). Section\,\ref{discuss} discusses the gas kinematics and provides a mass and energy budget for multiphase outflows. A summary is presented in Section\,\ref{summary}.

\section{Observations and data reduction}\label{data}
This section describes the observational datasets and reduction procedures: \textit{JWST}/MIRI-MRS mid-IR spectroscopy of the ionised and warm molecular gas, ALMA CO(2--1) observations tracing the cold molecular component, and ancillary \textit{Chandra} and \textit{Spitzer} data used for comparison.

\begin{figure*}[t]
\centering
\includegraphics[width = 1.28\columnwidth]{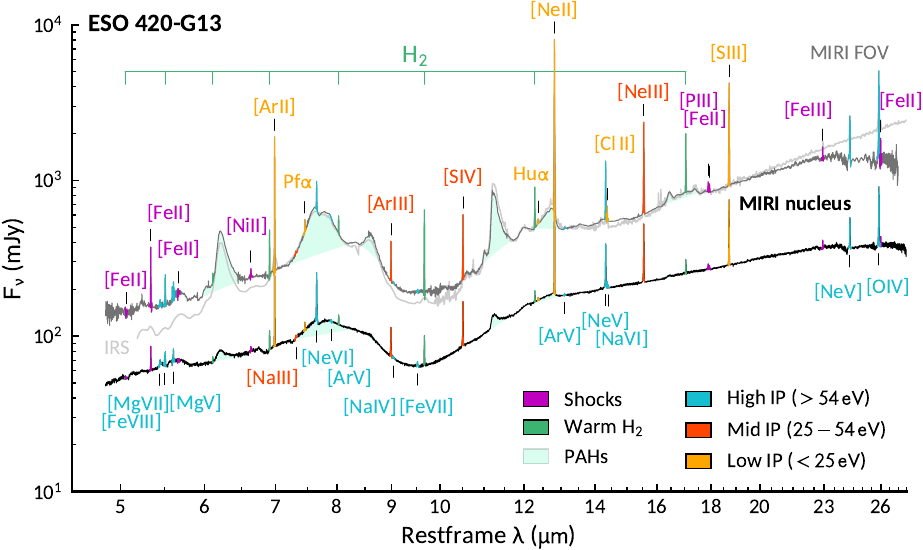}~
\includegraphics[width = 0.72\columnwidth]{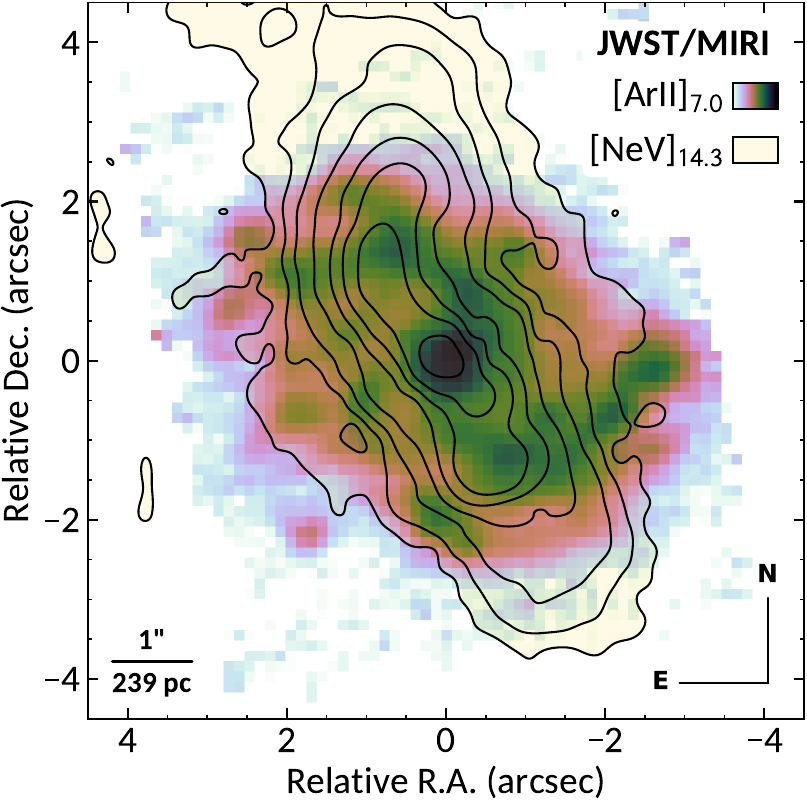}
\caption{\textit{Left:} \textit{JWST}/MIRI-MRS nuclear spectrum of ESO\,420-G13 (black line) extracted within a radius of $0\farcs7$, after subtracting the host-galaxy emission measured in an annulus between $0\farcs8$ and $1\farcs2$. The nucleus is about three times fainter than the total emission within the MIRI field of view (dark grey line), which agrees well with the \textit{Spitzer}/IRS spectrum from the \textsc{cassis} database \citep[light grey line;][]{lebouteiller15}. Emission lines are colour-coded according to their ionisation potential ($IP > 54\,\mathrm{eV}$, cyan; $25 < IP < 54\,\mathrm{eV}$, orange; $IP < 25\,\mathrm{eV}$, yellow), while shock-excited transitions and warm H$_2$ rotational lines are shown in violet and green, respectively. The main polycyclic aromatic hydrocarbon (PAH) bands are highlighted by light green shaded regions. \textit{Right:} [\ion{Ar}{ii}]$_{7.0}$ emission (background) traces a spiral distribution of star-forming regions, while [\ion{Ne}{v}]$_{14.3}$ contours reveal a highly collimated coronal structure. Contours are spaced by $0.3\, \mathrm{dex}$, starting at $1.4 \times 10^{-15}\,\mathrm{erg\,s^{-1}\,cm^{-2}}$. Axes show offsets relative to the nuclear ALMA 1.2\,mm continuum position ($\alpha$=04h\,13m\,49.69s, $\delta$=$-$32d\,00m\,25.129s; \citealt{jafo20}).}\label{fig_spec}
\end{figure*}

\subsection{MIRI/MRS spectroscopy}
ESO\,420-G13 was observed with the Medium-Resolution Spectroscopy (MRS) mode of the Mid-Infrared Instrument \citep[MIRI;][]{rieke15} onboard \textit{JWST} under programme GO\,1875 (PI: J.A. Fernández-Ontiveros). MIRI/MRS provides integral field spectroscopy over $5$--$28\,\mathrm{\micron}$ in four channels subdivided into three sub-bands each \citep{wells15}, with a FoV ranging from $3\farcs2 \times 3\farcs7$ (channel 1) to $6\farcs6 \times 7\farcs7$ (channel 4). Observations used a four-point dither pattern optimised for extended sources, with dedicated off-target exposures for thermal-background subtraction. Raw data from the Mikulski Archive (\href{https://doi.org/10.17909/5vc6-0j40}{DOI: 10.17909/5vc6-0j40}) were reduced with pipeline v1.12.5 through three stages: \textit{i)} ramp processing, including dark-current subtraction, bad-pixel and cosmic-ray removal, and linearity correction; \textit{ii)} spectral calibration \citep{labiano21}, including flat-fielding, astrometric and distortion corrections \citep{patapis24}, flux calibration \citep{law25}, and residual fringe correction \citep{argyriou20,crouzet25}; and \textit{iii)} cube reconstruction with background subtraction. Small sub-band flux discontinuities (at the level of a few percent) were corrected. No wiggle correction was applied, as no appreciable wiggles are apparent in the nuclear spectrum (Fig.\,\ref{fig_spec}).

For spatially resolved analysis, the continuum was removed per spaxel using a 1D polynomial fit, and moment maps were constructed for each transition. The noise was estimated from the median absolute deviation per spectral channel; only spaxels with emission $> 3\sigma$ over at least three consecutive channels were retained. The zeroth, first, and second moments provide integrated flux, line-of-sight velocity, and velocity dispersion, respectively, without correction for instrumental broadening in the latter.

The emission-line kinematics in the extracted spectra were measured by fitting each line with a single Gaussian plus a polynomial continuum using \texttt{lmfit}\footnote{\url{https://zenodo.org/records/16175987}}. Intrinsic velocity dispersions were obtained by subtracting the wavelength-dependent MIRI/MRS instrumental resolution \citep{argyriou23} in quadrature. The results are reported in Tables\,\ref{tab_h2kin} and \ref{tab_winds}.

\subsection{ALMA CO(2--1) observations}
Previous observations of the CO(2--1) transition at $230.54\, \mathrm{GHz}$ with ALMA for the central $3 \times 3\, \mathrm{kpc^2}$ in ESO\,420-G13 were obtained as part of the Twelve micron WInd STatistics (TWIST) project (Project ID: 2017.1.00236.S, P.I.: M.\,A. Malkan;  Fern\'andez-Ontiveros et al. in prep.). The aim of TWIST is to perform a systematic search for molecular gas outflows in the CO(2--1) line for 41 galaxies drawn from the $12$ micron sample \citep{rush93}. ESO\,420-G13 was observed with the ALMA 12\,m array on 8 December 2017 in band~6, targeting the CO(2--1) transition at a rest frequency of $230.5380\,\mathrm{GHz}$ ($T_\mathrm{ex} = 16.6\,\mathrm{K}$, $n_\mathrm{crit} = 2.7 \times 10^3\,\mathrm{cm^{-3}}$). The correlator was configured with a velocity resolution of $2.4\,\mathrm{\kms}$ over a total bandwidth of $2467\,\mathrm{\kms}$. Calibration followed the standard ALMA pipeline in \textsc{casa}~v5.1.1-5, and imaging was performed in \textsc{casa}~v5.4.0-70 using the \texttt{hogbom} deconvolution algorithm with \texttt{briggs} weighting (robustness parameter 2.0). The synthesised beam in the CO(2--1) spectral window is $0\farcs11 \times 0\farcs14$ (PA\,$=108\fdg5$), corresponding to $26.3 \times 33.5\,\mathrm{pc^2}$ at 49.4\,Mpc. The $27''$ ($\approx 6\,\mathrm{kpc}$) FoV was covered with a single pointing. Spectral cubes were produced with $0\farcs02$ pixels, $\sim 10\,\mathrm{km\,s^{-1}}$ channels, and corrected for primary beam attenuation. The final rms sensitivity is $0.02\,\mathrm{mJy\,beam^{-1}}$ in the continuum and $0.5\,\mathrm{mJy\,beam^{-1}}$ per $10\,\mathrm{\kms}$ channel. Further details of ALMA observations and reduction are given in \citet{jafo20}.

\subsection{Ancillary data}
A full-band ($0.5$--$10\,\mathrm{keV}$) image from \textit{Chandra}/ACIS-S observations \citep{lehmer10} was obtained from the \textit{Chandra} X-ray Center\footnote{\url{https://cxc.harvard.edu}} (Proposal ID: GO\,10393, PI: D.\,M. Alexander). For comparison with MIRI/MRS observations, we also include mid-IR spectra from the Combined Atlas of Sources with \textit{Spitzer}/IRS Spectra \citep[\textsc{cassis}][]{lebouteiller11,lebouteiller15}. The IRS spectrum in Fig.\,\ref{fig_spec} combines both the high- ($R = 600$; $10$--$29\, \mathrm{\micron}$) and low-spectral resolution modes ($R \sim 60$--$130$; $5.2$--$10\, \mathrm{\micron}$). The latter was scaled to match the $\sim 10\, \mathrm{\micron}$ continuum flux in the high-spectral resolution spectrum.


\section{Results}\label{results}

\begin{figure*}[!t]
  \centering
  \subfigure[]{\includegraphics[width = 0.97\columnwidth]{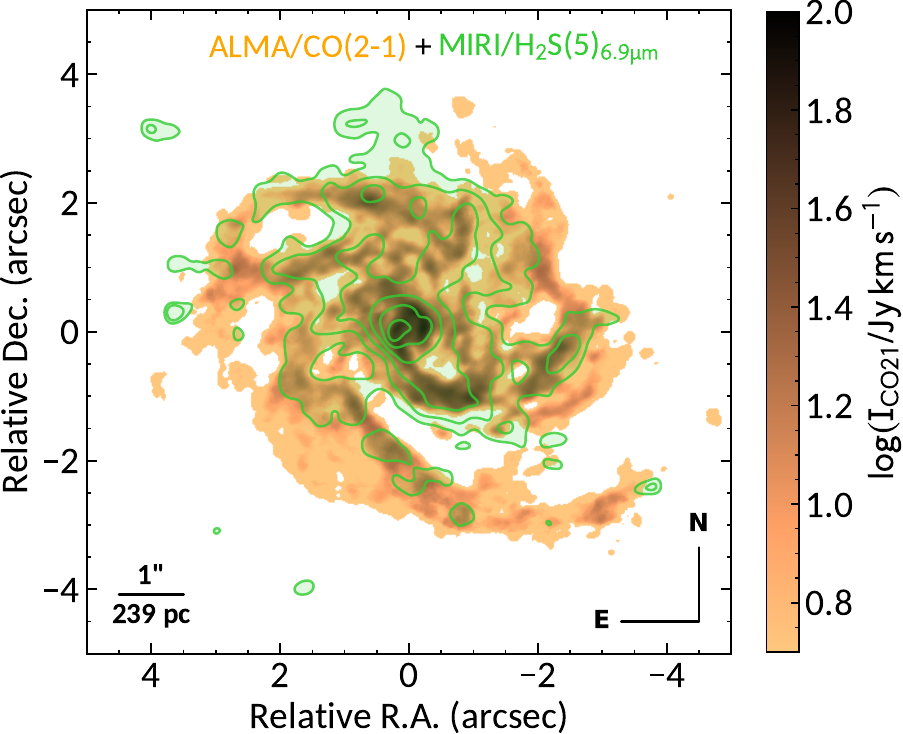}\label{subfig_COH2_flux}}~
  \subfigure[]{\includegraphics[width = \columnwidth]{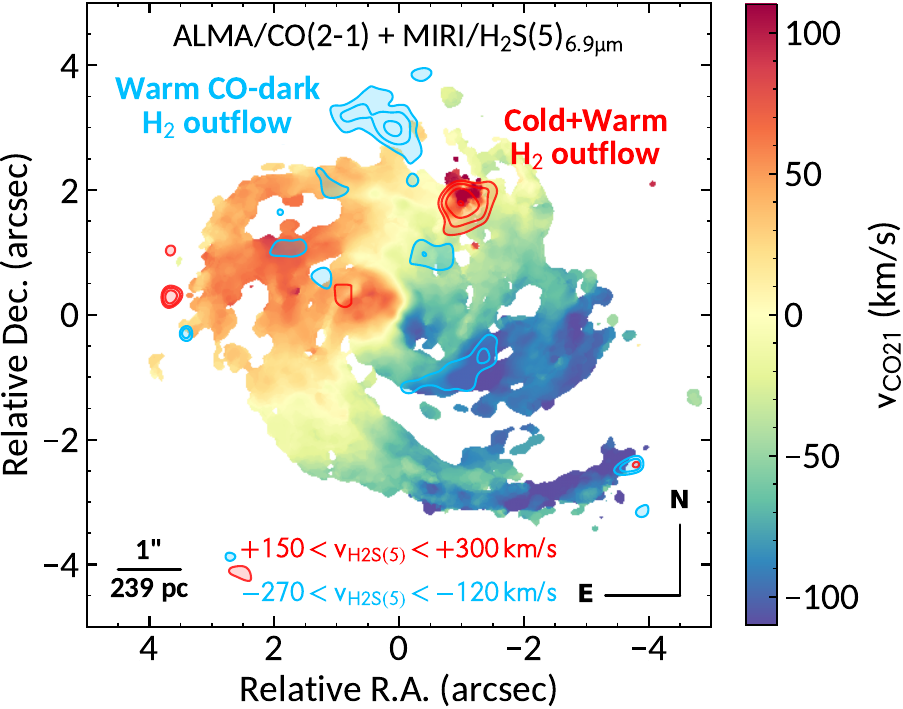}\label{subfig_COH2_vel}}\\[-0.1cm]
  \subfigure[]{\includegraphics[width = 0.97\columnwidth]{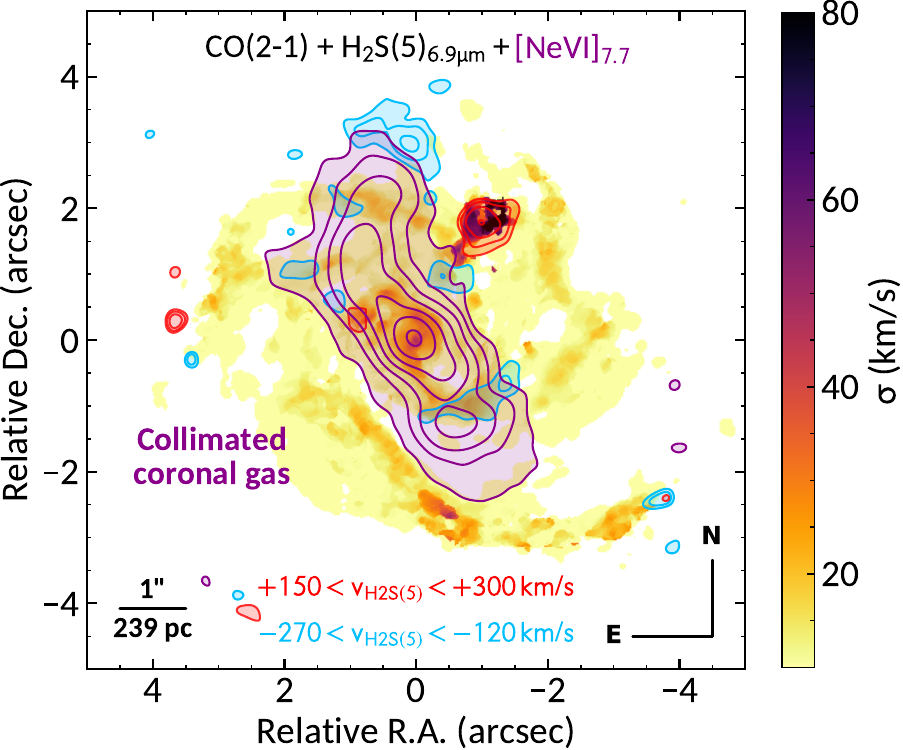}\label{subfig_COH2_sigma}}~
  \subfigure[]{\includegraphics[width = \columnwidth]{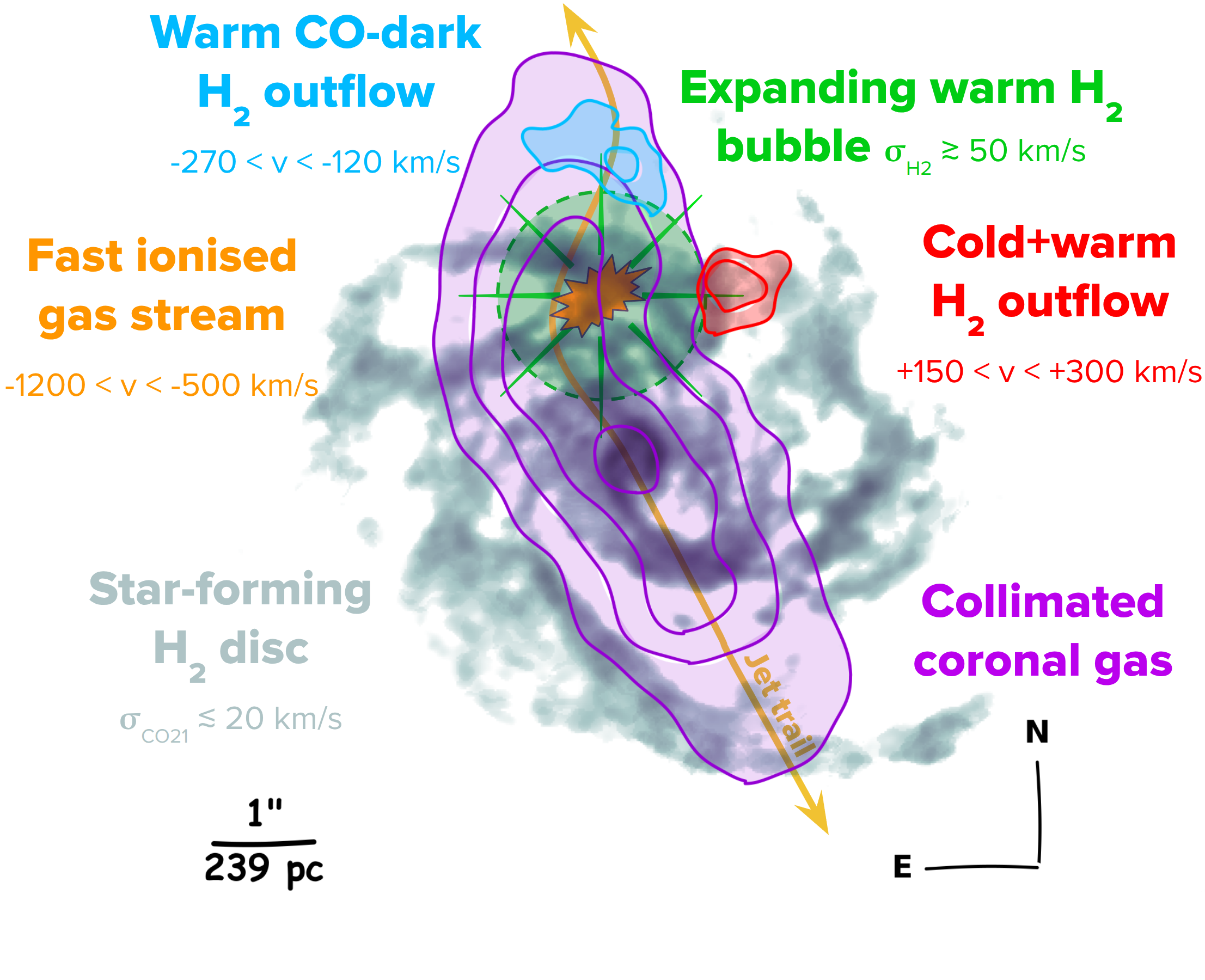}\label{subfig_COH2_sketch}}\\[-0.2cm]
  \caption{\textbf{(a)} ALMA CO(2--1) intensity map compared to \textit{JWST}/MIRI warm H$_2$\,S(5) emission at $6.9\, \mathrm{\micron}$ (green contours). Note the extended warm H$_2$ emission at $\Delta \mathrm{Dec.} > 2\farcs5$ to the north ($\gtrsim 600\, \mathrm{pc}$), with no CO(2--1) counterpart. \textbf{(b)} CO(2--1) average velocity map compared to the blueshifted ($-270$ to $-120\, \mathrm{\kms}$; blue contours) and redshifted ($150$ to $300\, \mathrm{\kms}$; red contours) warm H$_2$\,S(5) outflows. \textbf{(c)} CO(2--1) average velocity dispersion map compared to the warm H$_2$ outflows and the collimated [\ion{Ne}{vi}]$_{7.7}$ coronal gas emission (purple contours). \textbf{(d)} Schematic representation of the different components in the centre of ESO\,420-G13.}\label{fig_COH2}
\end{figure*}

\begin{figure*}[!t]
  \centering
  \subfigure[]{\includegraphics[width = \columnwidth]{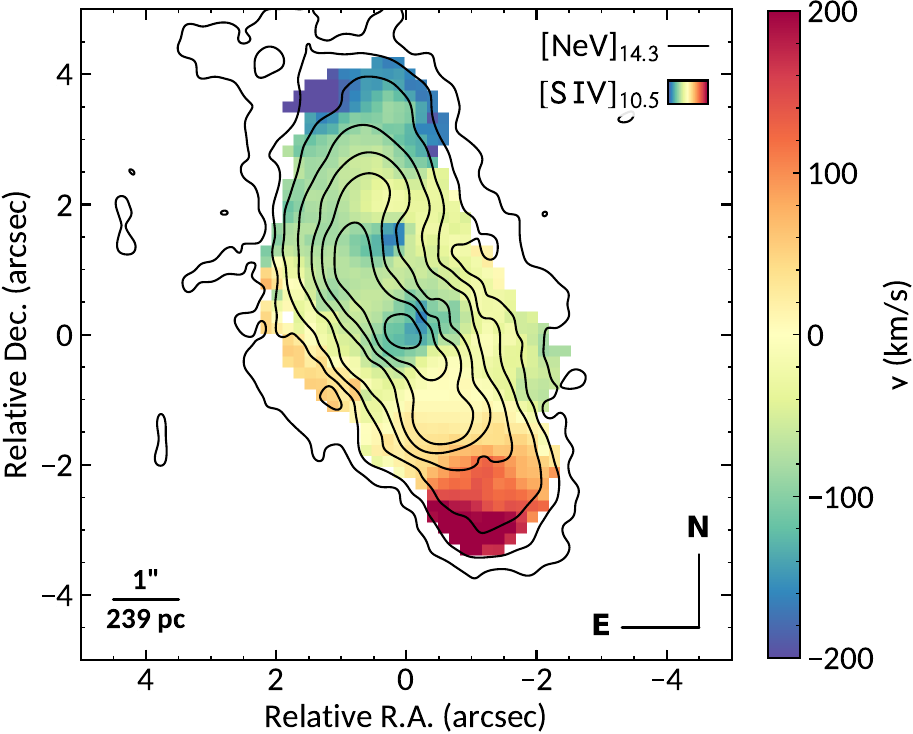}\label{subfig_s4_vel}}~
  \subfigure[]{\includegraphics[width = \columnwidth]{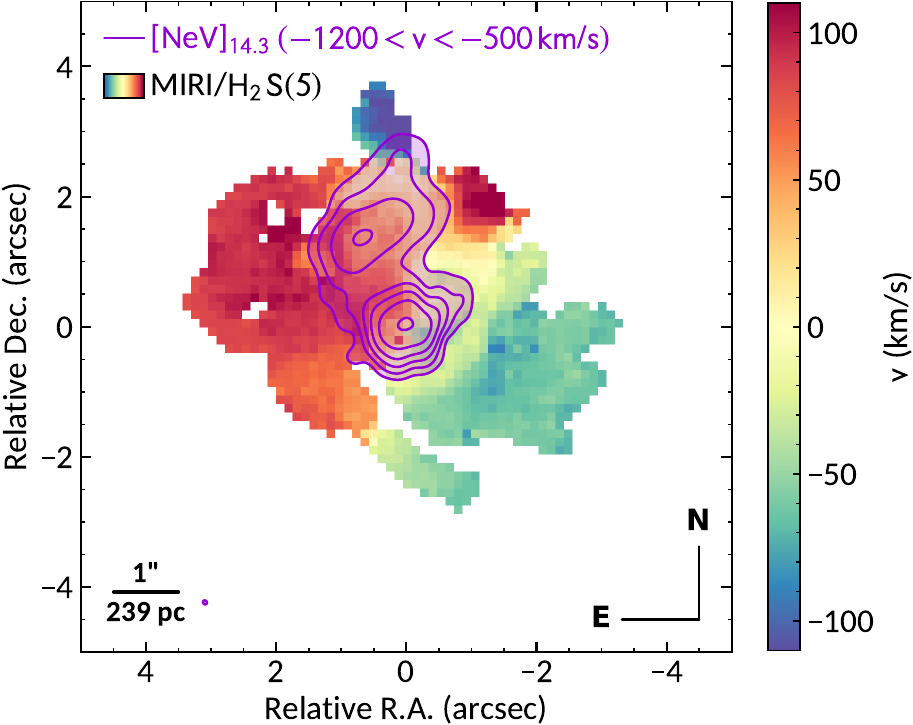}\label{subfig_H2_vel}}\\[-0.3cm]
  \subfigure[]{\includegraphics[width = \columnwidth]{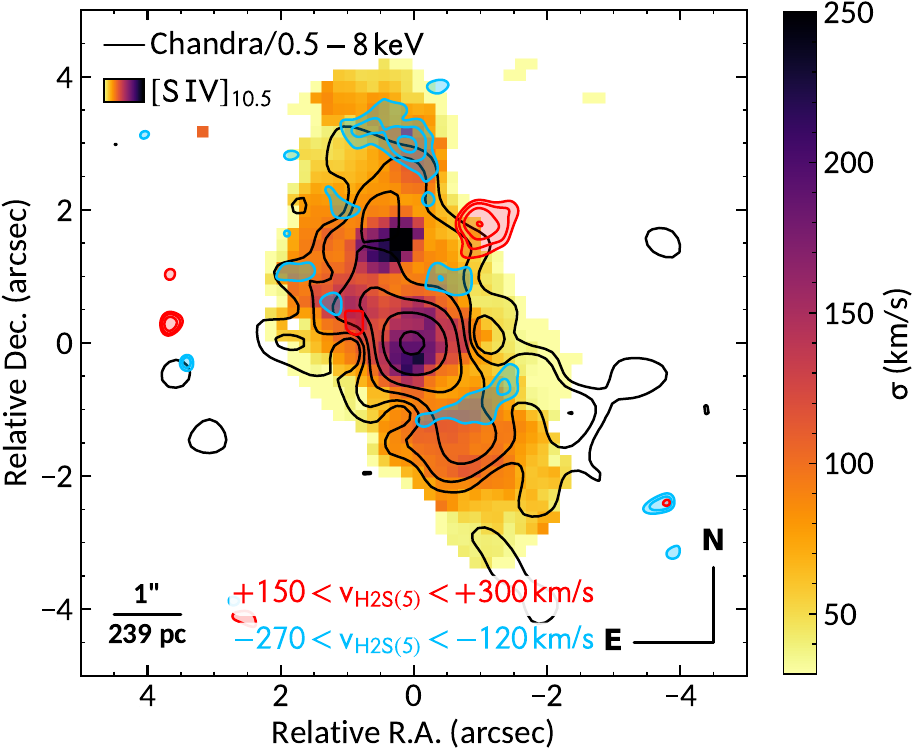}\label{subfig_s4_sig}}~
  \subfigure[]{\includegraphics[width = \columnwidth]{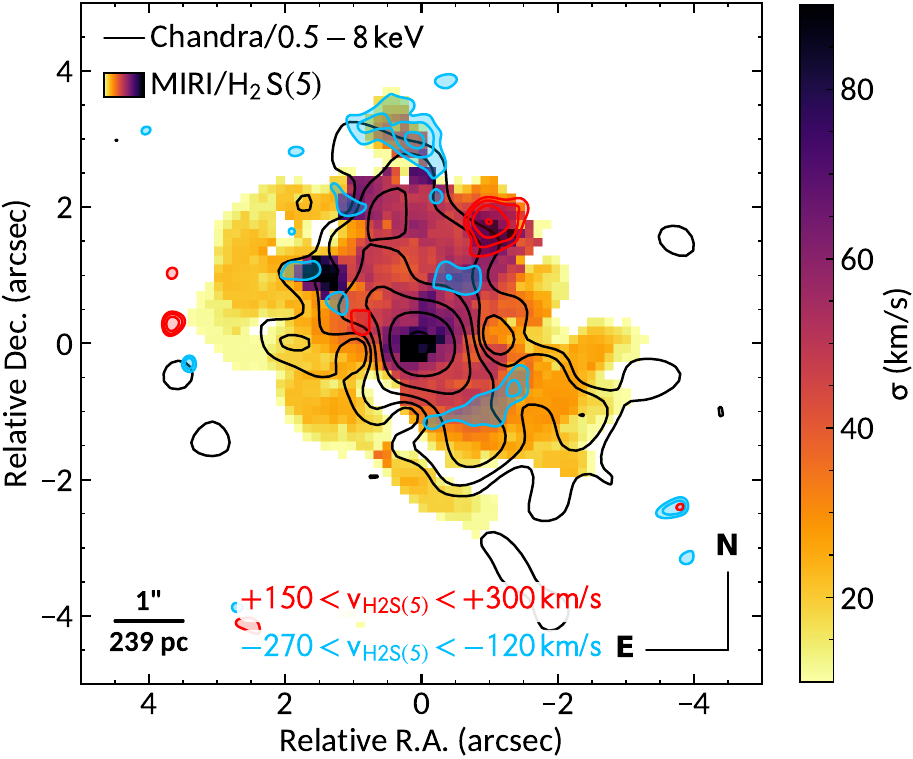}\label{subfig_H2_sig}}\\[-0.2cm]
  \caption{Integrated [\ion{Ne}{v}]$_{14.3}$ line emission (black contours), and the fast ionised gas stream in [\ion{Ne}{v}]$_{14.3}$ ($-1200$ to $-500\, \mathrm{\kms}$) (purple contours), are shown over the [\ion{S}{iv}]$_{10.5}$ and H$_2$\,S(5) average velocity maps in panels \textbf{(a)} and \textbf{(b)}, respectively. Contours of the \textit{Chandra}/$0.5$--$8\, \mathrm{keV}$ emission (in black) and the warm molecular gas outflows detected in H$_2$\,S(5) at $6.9\, \mathrm{\micron}$ (in blue and red) are compared to the average velocity dispersion maps of the [\ion{S}{iv}]$_{10.5}$ \textbf{(c)} and the H$_2$\,S(5) transitions \textbf{(d)}. The highly-ionised gas shows negative velocities ($\lesssim -50\, \mathrm{\kms}$) along the north-west direction in the collimated gas plume (a), in contrast with the receding velocities seen in the warm H$_2$ disc (b). Blueshifted coronal gas at $\sim 1\farcs5$ north of the nucleus (b) traces broadened line emission ($\sigma_{[SIV]} \gtrsim 200\, \mathrm{\kms}$), elongated along a perpendicular direction to the major axis of the coronal gas emission (c), suggesting a possible bend of the jet trail at this point. Large $\sigma_{[SIV]}$ values $> 100\, \mathrm{\kms}$ are also detected $\sim 2''$ south of the nucleus, and at the top of the northern plume (c). The warm H$_2$ gas surrounding this fast ionised gas stream shows enhanced turbulence ($\sigma_{H_2S(5)} \gtrsim 50\, \mathrm{\kms}$; d), in contrast with the cold molecular gas ($\sigma_{CO21} < 20\, \mathrm{\kms}$; Fig.\,\ref{subfig_COH2_sigma}).}\label{fig_velsigma}
\end{figure*}

The resulting MIRI/MRS spectra for the nucleus (black) and the total FoV (dark grey) of ESO\,420-G13 are presented in Fig.\,\ref{fig_spec}, extracted with \textsc{photutils} \citep{bradley24}. The total FoV spectrum was obtained from the largest area common to all four MIRI/MRS channels, and is thus limited by the smaller FoV of Channel~1. It is in excellent agreement with previous \textit{Spitzer}/IRS observations (light grey) across most of the wavelength range. The main discrepancies occur above $\sim 20\,\mathrm{\micron}$, where the MIRI sensitivity decreases due to the warm mirror, and below $\sim 7\,\mathrm{\micron}$ as well as around $\sim 10\,\mathrm{\micron}$, where the lower \textit{Spitzer} fluxes are likely due to a slight over-subtraction of the background emission. Nevertheless, the mismatch at long wavelengths does not have a systematic effect on the measured emission-line fluxes: the MIRI [\ion{Ne}{v}]$_{24.3}$ and [\ion{O}{iv}]$_{25.9}$ line fluxes differ by $+6\%$ and $-10\%$, respectively, from previous \textit{Spitzer} measurements (Table\,\ref{tab_ionflux}; \citealt{jafo16}). This effect becomes more noticeable only at very low continuum flux levels ($\lesssim 200\,\mathrm{mJy}$). No spectral leak is detected at $\sim 12.2\,\mathrm{\micron}$, a typical feature seen in spectra with very steep red continua\footnote{\url{https://jwst-pipeline.readthedocs.io/en/stable/jwst/spectral_leak/main.html}}. After subtracting the underlying host-galaxy contribution, the nuclear spectrum shows a significant decrease in PAH strengths at $6.2$, $7.7$, $8.6$, $11.3$, $12.0$, and $17.0\,\mathrm{\micron}$ compared to integrated MIRI FoV emission.

\subsection{A CO-dark outflow revealed by warm molecular gas}\label{res_H2}
The ALMA CO(2--1) observations revealed a massive molecular gas outflow ($M_\text{H2} \sim 8.3 \times 10^{6}\, \mathrm{M_\odot}$) with velocities of $+150$ to $+300\, \mathrm{km\,s^{-1}}$ in ESO\,420-G13, located $340$--$600\,\mathrm{pc}$ north-west of the active nucleus \citep{jafo20}. No blueshifted counterpart was detected to a mass limit of $\sim 10^5\,\mathrm{M_\odot}$. Based on the collimated morphology and the energy and momentum budgets, we ruled out possible launching mechanisms such as AGN radiation pressure, star-formation winds, and supernovae, favouring instead a scenario in which the outflow is driven by a previously undetected compact jet in this galaxy. ESO\,420-G13 is the second known case, after NGC\,1377, where a previously unknown jet has been identified through its interaction with the ISM \citep{aalto12}. This result suggests that, if such radiatively faint jets are common in low-luminosity AGNs, their kinetic power --\,and hence their feedback potential\,-- may remain largely overlooked in the absence of observations capable of detecting jet--ISM interactions.

Figure\,\ref{subfig_COH2_flux} compares the cold molecular gas distribution with the warm H$_2$ emission traced by the 0--0 S(5) transition at 6.9\,$\mu$m. The H$_2$\,S(5) line has lower and upper energy levels of $3\,475\,\mathrm{K}$ and $4\,586\,\mathrm{K}$, respectively \citep{roueff19}, and was selected as the brightest H$_2$ feature detected in MIRI/MRS channel~1, which offers the best spectral and spatial resolution. The green contours reveal an overall morphological correspondence between the warm and cold molecular gas, except for an extended plume of warm H$_2$ located at $\Delta\alpha \sim 0''$, $\Delta\delta \sim +3''$, about $\sim 720\,\mathrm{pc}$ north of the nucleus. This region is CO-dark, that is, it lacks a CO(2--1) counterpart, and its kinematics show an average blueshifted velocity of $-30\,\mathrm{\kms}$, reaching terminal velocities of $\sim -270\,\mathrm{\kms}$ (see the blue contours in Figs.\,\ref{subfig_COH2_vel} and \ref{subfig_COH2_sigma}). We also confirm the presence of warm molecular gas associated with the redshifted outflow, with spatial and velocity distributions closely matching those of the CO(2--1) counterpart (Fig.\,\ref{subfig_COH2_vel}).

While asymmetric outflows are not uncommon \citep[e.g.][]{lutz20}, the presence of spatially separated blueshifted CO-dark and redshifted CO-bright components in ESO\,420-G13, both emerging north of the nucleus, points to a more complex jet--ISM interaction than the symmetric bipolar picture often assumed in AGNs. This geometry also revisits the `missing' blueshifted CO counterpart discussed in our previous ALMA analysis, where we considered jet--ISM configurations that could naturally produce a one-sided outflow, such as a jet impacting a discrete molecular cloud (or relic material) present only on one side, or a jet whose trajectory bends towards the disc on one side, thereby producing a single dominant impact region \citep{jafo20}. In the following, we use high-ionisation lines to trace the gas excitation produced by the jet along its path and compare its morphology with the warm and cold molecular gas outflows.

\begin{figure*}[!t]
\centering
\subfigure[]{\includegraphics[width = \columnwidth]{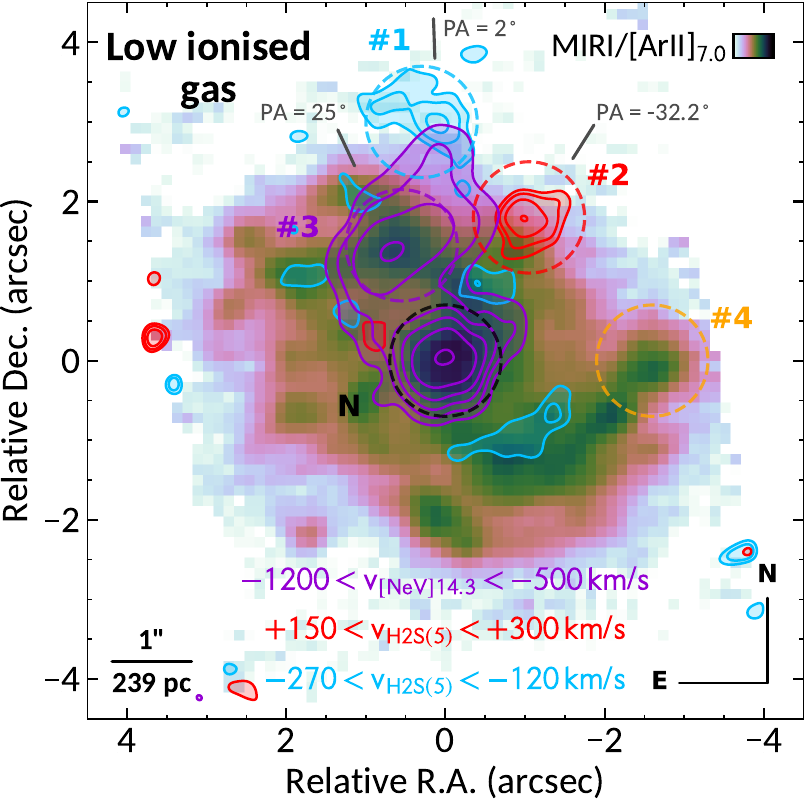}\label{subfig_hii}}~
\subfigure[]{\includegraphics[width = \columnwidth]{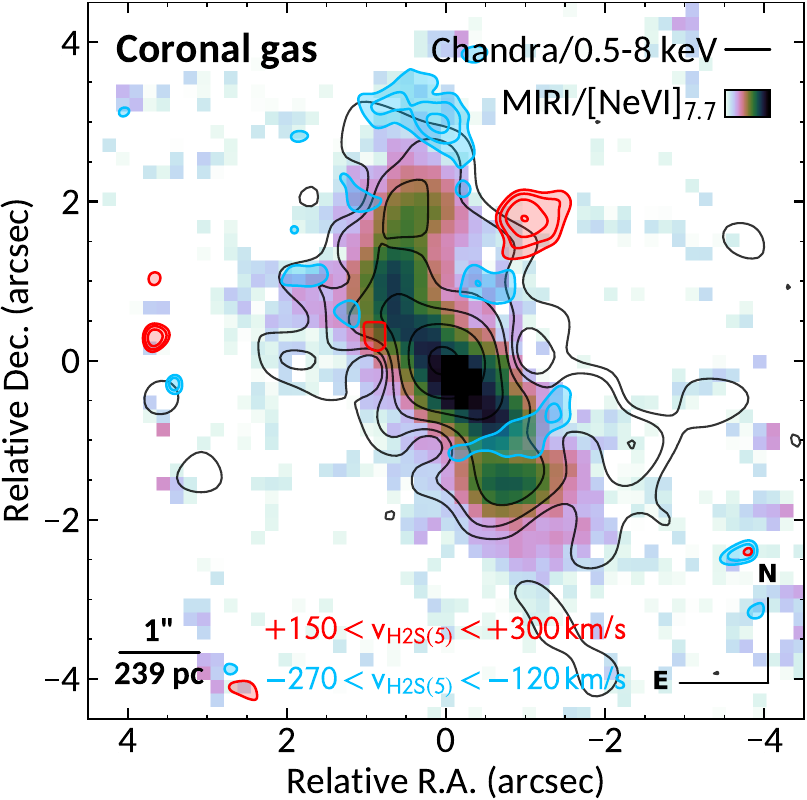}\label{subfig_jet}}\\[-0.2cm]
\caption{\textbf{(a)} The background image shows the [\ion{Ar}{ii}]$_{7.0}$ emission line map, tracing the star-forming regions in the central disc of ESO\,420-G13. The contours indicate the outflows detected in the warm H$_2$\,S(5) emission at $6.9\, \mathrm{\micron}$ coloured according to their relative velocity (in blue for $-270 < v < -120\, \mathrm{\kms}$, and red for $+150 < v < +300\, \mathrm{\kms}$), and the fast ionised gas stream detected in the [\ion{Ne}{v}]$_{14.3}$ emission line (in purple, $-1200 < v < -500\, \mathrm{\kms}$). The orientations of the three pseudo-slits used for the position-velocity diagrams in Fig.\,\ref{fig_pvmaps} are indicated by short grey lines. \textbf{(b)} The collimated [\ion{Ne}{vi}]$_{7.7}$ emission-line map shows a remarkably similar morphology when compared to the \textit{Chandra} $0.5$--$8\, \mathrm{keV}$ continuum distribution (black contours; \citealt{lehmer10}), suggesting a common excitation mechanism for the origin of the X-ray emission and the coronal gas. Additionally, we also show the contours for the warm H$_2$\,S(5) outflows (in blue and red).}\label{fig_regions}
\end{figure*}

\subsection{Collimated coronal gas extended emission}\label{res_coronal}
The direct impact of the jet on the ISM of ESO\,420-G13 is revealed by emission lines from mid- to high-ionisation gas ($IP \gtrsim 25\,\mathrm{eV}$), particularly coronal lines ($\gtrsim 100\,\mathrm{eV}$), which require extremely energetic processes to form. The coronal gas traced by the [\ion{Ne}{v}]$_{14.3}$ and [\ion{Ne}{vi}]$_{7.7}$ transitions ($97.12$ and $126.2\,\mathrm{eV}$, respectively) exhibits a collimated morphology along the north-east--south-west direction (Figs.\,\ref{subfig_COH2_sigma} and \ref{subfig_s4_vel}). At $\Delta\alpha \sim +1\farcs4$, $\Delta\delta \sim +0\farcs7$ relative to the nucleus, the orientation of the extended emission bends northwards, while the southern end shows a slight twist to the south. The morphology of the blueshifted warm H$_2$ outflow closely follows the outer boundary of the coronal gas plume. This spatial correspondence suggests a possible interaction between the two components, supporting the scenario in which the jet drives the warm molecular outflow (see Section\,\ref{disc_bubble}). A schematic representation of the jet--ISM interaction in the centre of ESO\,420-G13 and the different components involved is shown in Fig.\,\ref{subfig_COH2_sketch}.

The decoupling between the motions of the highly ionised gas and the disc rotation is evident from the kinematic analysis. The north-eastern part of the jet, projected over the receding side of the disc, is blueshifted ($\sim -50$ to $-150\,\kms$), while the south-western jet is redshifted ($0$ to $+200\,\kms$) over the approaching side of the disc (Figs.\,\ref{subfig_s4_vel} and \ref{subfig_H2_vel}). Similar morphology and kinematics to those observed in [\ion{Ne}{v}]$_{14.3}$ and [\ion{Ne}{vi}]$_{7.7}$ are also seen in other mid- to high-ionisation lines, namely [\ion{S}{iv}]$_{10.5}$, [\ion{Ne}{v}]$_{24.3}$, and [\ion{O}{iv}]$_{25.9}$. Weaker high-ionisation lines such as [\ion{Ar}{v}]$_{7.9,13.1}$, [\ion{Mg}{v}]$_{5.6}$, [\ion{Mg}{vii}]$_{5.5}$, [\ion{Fe}{vii}]$_{9.5}$, [\ion{Fe}{viii}]$_{5.4}$, [\ion{Na}{iii}]$_{7.3}$, and [\ion{Na}{iv}]$_{9.0,14.4}$ display similar trends, although their collimated emission is less extended, likely in part due to their lower S/N.

The morphology and kinematics of the low-ionisation gas, in contrast to the high-ionisation gas, follow those of the rotating molecular gas disc. This is illustrated, for example, by the [\ion{Ar}{ii}]$_{7.0}$ line map in Fig.\,\ref{subfig_hii}, which traces the spiral arms and the distribution of star-forming clusters also seen in the CO(2--1) line. Particularly notable is the absence of hydrogen recombination emission, for example Pf$\alpha$, suggesting over-ionisation within the collimated coronal emission region (see Section\,\ref{disc_coronal}). Interestingly, mid-ionisation lines such as [\ion{Ne}{iii}]$_{15.6}$ and [\ion{Ar}{iii}]$_{9.0}$ display mixed characteristics (not shown): the blue half of these lines contains both the blueshifted north-eastern jet and the approaching south-western disc, while the red half traces the opposite configuration, i.e. the south-western jet and the receding north-eastern disc. We also note that the $1''$ north-westwards extension of [\ion{Ne}{ii}]$_{12.8}$ seen in the VISIR narrow-band image \citep{jafo20} is consistent with the MIRI/MRS map. However, this structure does not correspond to the jet, as previously suggested in \citet{jafo20}, but rather to the inner tip of the northern star-forming arm (see Fig.\,\ref{subfig_hii}).

Overall, the highly ionised gas in ESO\,420-G13 traces a collimated structure associated with the jet, whereas the low-ionisation gas largely follows the rotating molecular disc and the star-forming spiral arms. Motivated by previous studies of jet--ISM interactions in nearby Seyfert galaxies \citep{rodriguez17,may18}, we compared the coronal gas distribution with the morphology of the X-ray emission. ESO\,420-G13 is a weak X-ray source ($L_\mathrm{2-10keV} \sim 10^{40}\,\mathrm{\ergs}$), and earlier studies favoured a non-AGN origin for its high-energy emission \citep{lehmer10,torres-alba18}. Nevertheless, the $0.5$--$8\,\mathrm{keV}$ continuum observed by \textit{Chandra} is extended and closely matches the morphology of the coronal gas emission revealed by MIRI/MRS (Fig.\,\ref{subfig_jet}). This spatial correspondence suggests that the X-ray emission is indeed associated with jet activity, as seen in other Seyfert nuclei \citep{bianchi07,wang12,rodriguez17,may18,trindade23}. In particular, numerical simulations suggest that mixing of hot coronal gas from the jet with cold ISM clouds may lead to significant enhancement of bremsstrahlung cooling, producing the observed X-ray emission \citep{nims15,ward26}. The collimated emission in the nucleus of ESO\,420-G13 is therefore interpreted as gas ionised by the jet along its path. In this scenario, shocks initially heat the gas as the jet propagates through the ISM, creating a hot coronal phase. The subsequent mixing of this shock-heated coronal gas with cold clouds then produces efficient bremsstrahlung cooling, generating both the extended soft X-ray continuum and enhanced coronal line emission. This interpretation is consistent with jet-driven ionisation scenarios observed in other active galaxies \citep{binette85,bicknell98,wilson99,wang12,fabbiano17,fabbiano22}.

\begin{figure*}
\centering
\includegraphics[width = 2\columnwidth]{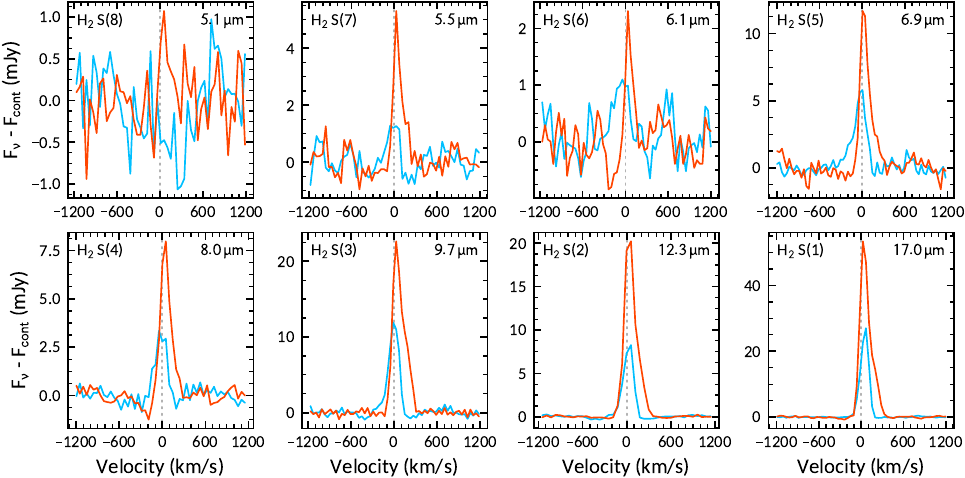}
\caption{Continuum-subtracted rotational H$_2$ transitions from S(8) to S(1) observed by MIRI in regions \#1 (blue) and \#2 (red). S(2) to S(7) lines in region \#1 are blueshifted relative to region \#2, with more energetic transitions exhibiting increasingly larger velocity shifts.}\label{fig_H2spec}
\end{figure*}

\subsection{A fast ionised gas stream}\label{res_stream}
A high-velocity gas stream ($\sim -1200$ to $-500\,\kms$) is detected in several mid- to high-ionisation lines --\,[\ion{S}{iv}]$_{10.5}$, [\ion{Ne}{iii}]$_{15.6}$, [\ion{Ne}{v}]$_{14.3,24.3}$, [\ion{Ne}{vi}]$_{7.7}$, and [\ion{O}{iv}]$_{25.9}$\,-- at the location where the jet bends northwards ($\Delta\alpha \sim +1\farcs4$, $\Delta\delta \sim +0\farcs7$). This component is spatially extended on the northern side of the jet, elongated towards the midpoint between the blue- and redshifted warm H$_2$ outflows (Fig.\,\ref{subfig_hii}). This gas stream is also evident in the average observed velocity dispersion map of the [\ion{S}{iv}]$_{10.5}$ line, whereas the warm H$_2$ gas surrounding this structure shows very high velocity dispersion values of $\gtrsim 50$--$100\,\kms$ (Figs.\,\ref{subfig_s4_sig} and \ref{subfig_H2_sig}), in contrast to the low dispersion measured in the CO(2--1) cold molecular gas ($\lesssim 20\,\kms$). Given the $\sim 10\, \mathrm{\kms}$ absolute velocity accuracy of MIRI/MRS at the wavelength of the H$_2$ transitions \citep{argyriou23}, the observed variations across the H$_2$ velocity-dispersion map are significant. This contrast suggests that enhanced turbulence in the coronal gas, which occurs where the jet bends its path, is likely driving the expansion of a warm molecular gas bubble traced by the H$_2$\,S(5) velocity-dispersion map \citep{mukherjee18b}. In contrast, most of the cold molecular gas appears to be unaffected by the jet, except for the redshifted outflow.

To further investigate the properties of the fast ionised gas stream and its relation to molecular gas outflows, we extracted spectra from five representative regions sampling the nucleus, the two warm H$_2$ outflows, the fast ionised gas stream, and a star-forming region in the disc (Fig.\,\ref{subfig_hii}). The latter was selected as a reference for comparison because it does not overlap with the coronal gas emission, unlike most of the other star-forming knots in the nuclear disc. In all cases, we used an aperture radius of $0\farcs7$, corresponding to 5 and 2 resolution elements in MIRI/MRS channels~1 and 4, respectively. Line fluxes for all the ionic and H$_2$ rotational transitions detected in the spectra are listed in Appendix\,\ref{app_fluxes}.

The spectrum extracted at the location of the fast ionised gas stream (Fig.\,\ref{fig_windspec}) is characterised by very strong emission from high-ionisation lines: for instance, [\ion{O}{iv}]$_{25.9}$ is the second brightest feature after [\ion{Ne}{ii}]$_{12.8}$ and is stronger than [\ion{S}{iii}]$_{18.7}$, while [\ion{Ne}{v}]$_{24.3}$ has a flux comparable to [\ion{Ne}{iii}]$_{15.6}$. This indicates enhanced gas excitation compared to the nucleus. The high-velocity broad component associated with the gas stream is $\gtrsim 5$ times fainter than the corresponding narrow components and is not detected in any of the weaker coronal lines (e.g. [\ion{Mg}{v}]$_{5.6}$), likely due to the lower S/N ratio. It is only marginally detected in some of the brightest low-ionisation lines (e.g. [\ion{S}{iii}]$_{18.7}$). The weakness of the blueshifted broad component in the low-ionisation transitions suggests that the fast ionised gas stream is composed of highly excited and possibly matter-bounded gas, consistent with shock-ionised material produced by the interaction of the jet with the surrounding medium.


\section{Discussion}\label{discuss}
In this section, we discuss the physical interpretation of the multiphase gas structures identified in Section\,\ref{results}. We first examine whether the warm molecular gas kinematics are consistent with an expanding H$_2$ bubble driven by the jet--ISM interaction, and then estimate the mass, momentum, and energy budget of the molecular and ionised outflows.

\subsection[An expanding warm H2 bubble?]{An expanding warm H$_2$ bubble?}\label{disc_bubble}
In ESO\,420-G13, the warm H$_2$ emission is not co-spatial with the jet trail, but instead avoids the coronal gas region, in contrast to other Seyfert nuclei with jet-driven outflows such as IC\,5063 \citep{dasyra15,dasyra24}, NGC\,4258 \citep{ogle14}, NGC\,1386 \citep{mezcua15,rodriguez17}, ESO\,428-G14 \citep{may18}, and NGC\,7319 \citep{ps22}. Most of the warm H$_2$ gas co-rotates with the cold molecular component in the nuclear star-forming disc, while regions of enhanced turbulence and two warm H$_2$ outflows surround the coronal gas and the fast ionised gas stream described in Section\,\ref{res_stream}. To characterise the physical properties of the warm molecular gas, we constructed rotational diagrams (Boltzmann plots) using the measured H$_2$ line fluxes for the S(1)--S(8) transitions within the $0\farcs7$ apertures defined in Fig.\,\ref{subfig_hii}. These diagnostics represent the column density ($N_\text{u}$) of each transition --\,normalised by the statistical weight of the upper level ($g_\text{u}$)\,-- as a function of the corresponding upper-level energy ($E_\text{u}$). The normalised column densities were calculated from the measured line fluxes in Table\,\ref{tab_h2flux} using the Einstein coefficients from \citet{roueff19}. Appendix\,\ref{app_H2} illustrates the resulting rotational diagrams for the five regions shown in Fig.\,\ref{subfig_hii}. The excitation temperature and the warm molecular gas mass were derived by fitting $N_\text{u} / g_\text{u}$ versus $E_\text{u}$ \citep[e.g.][]{rigopoulou02}. In all regions, the different slopes traced by the low-J and high-J transitions indicate that a single-temperature component cannot reproduce the observed distribution. For these calculations, we adopted an ortho-to-para ratio of 3, consistent with local thermodynamic equilibrium (LTE) at $T \gtrsim 200\,\mathrm{K}$ \citep{burton92,sheffer11}.

The excitation temperatures and masses obtained for each region are indicated in the corresponding panels of Fig.\,\ref{fig_H2}. The redshifted outflow (\#2) contains the largest warm molecular gas mass, $M^\mathrm{warm}_\mathrm{H_2} \sim 3.9 \times 10^5\,\mathrm{M_\odot}$. In all regions, the warm H$_2$ mass is dominated by the cooler component ($290$--$470\,\mathrm{K}$), while the hotter component ($890$--$1900\,\mathrm{K}$) contributes only a small fraction. The redshifted H$_2$ outflow (\#2) is twice as massive as the blueshifted outflow (\#1), which contains $1.8 \times 10^5\,\mathrm{M_\odot}$. The highest temperatures ($\sim 470\,\mathrm{K}$ and $1900\,\mathrm{K}$) are found in the fast ionised gas stream (\#3), which also has the lowest warm molecular mass ($10^5\,\mathrm{M_\odot}$). Overall, the warm H$_2$ temperatures are comparable to those measured in radio galaxies \citep{ogle10,ps22,ogle24,dasyra24}, where the heating is consistent with kinetic energy dissipation from radio jets via shocks and/or cosmic rays. In contrast, the star-forming region (\#4) shows the lowest excitation temperatures ($\sim 290\,\mathrm{K}$ and $890\,\mathrm{K}$), comparable to those derived for the integrated MIRI/MRS FoV. Moreover, the H$_2$\,S(3)/PAH$_{11.3}$ ratio is $\sim 0.01$ both in region~\#4 and in the integrated FoV, consistent with values typically found in star-forming environments \citep{riffel25}. This suggests that star formation dominates the warm H$_2$ emission over most of the central kiloparsec, although the enhanced velocity dispersion and temperatures around the expanding bubble indicate an additional contribution from shock excitation associated with the jet--ISM interaction \citep{lopez25}. The total warm H$_2$ mass derived for the MIRI/MRS FoV is $1.2 \times 10^7\,\mathrm{M_\odot}$, approximately 25 times smaller than the total cold molecular gas mass of $\sim 3 \times 10^8\,\mathrm{M_\odot}$ estimated from the ALMA observations \citep{jafo20}.

A $3\sigma$ upper limit of $\lesssim$$10^5\, \mathrm{M_\odot}$ for the cold molecular gas mass in the blueshifted outflow can be derived from the ALMA observations, adopting a CO excitation ratio of $r_{21} = L'_\mathrm{CO(2-1)}/L'_\mathrm{CO(1-0)} \sim 0.7$ \citep{leroy22} and a CO-to-H$_2$ conversion factor of $\alpha_\mathrm{CO} = 0.8\,\mathrm{M_\odot\,(K\,\kms\,pc^2)^{-1}}$, typical of luminous and ultra-luminous IR galaxies \citep[LIRGs and ULIRGs;][]{bolatto13}. This contrasts with the $M^\mathrm{cold}_\mathrm{H2} = 8.3 \pm 0.7 \times 10^6\,\mathrm{M_\odot}$ measured for the redshifted outflow \citep{jafo20}. An estimate of the CO-dark molecular mass can be obtained from the cold-to-warm H$_2$ mass ratio, $M^\mathrm{cold}_\mathrm{H2}/M^\mathrm{warm}_\mathrm{H2} \sim 21$, derived for the redshifted outflow. Assuming a similar ratio for the blueshifted outflow yields a cold molecular mass of $\sim 3.8 \times 10^6\,\mathrm{M_\odot}$, approximately an order of magnitude above the ALMA detection limit. These masses would increase by a factor of $5.4$ if a Milky Way conversion factor of $\alpha_\mathrm{CO} = 4.3\,\mathrm{M_\odot\,(K\,\kms\,pc^2)^{-1}}$ is adopted \citep{bolatto13}, and decrease by a factor of $\sim 3$ if the CO emission is optically thin \citep{combes13,dasyra16}. An independent estimate based on the extrapolation of the H$_2$ Boltzmann diagram down to a lower temperature cutoff of 50\,K \citep{togi16} yields a total H$_2$ mass of $3.8 \times 10^7\,\mathrm{M_\odot}$, an order of magnitude larger.

The absence of CO(2--1) emission in the blueshifted outflow may reflect either CO destruction by photodissociative shocks and/or cosmic rays generated by the jet \citep{neufeld89,papadopoulos18}, or alternatively, a shift of the CO excitation towards mid- to high-J transitions, where CO(2--1) emits inefficiently. In the case of cosmic-ray dissociation, the required reduction in the CO-to-H$_2$ abundance implies ionisation rates $\gtrsim 30$ times higher than the Galactic value \citep{bisbas17}, consistent with strong particle acceleration in jet--ISM interactions. On the other hand, the CO spectral line energy distributions (SLEDs) of luminous AGNs typically peak between CO(3--2) and CO(7--6), with CO(2--1) fluxes representing $\sim$20--60\% of the peak transition \citep[e.g.][]{li24}. The uncertainty in this ratio could increase the ALMA upper mass limit of $10^5\, \mathrm{M_\odot}$ by up to a factor of $\sim$3, still an order of magnitude below the cold molecular mass estimated for the redshifted outflow. A much lower CO(2--1) fraction would require extreme X-ray dominated region (XDR) conditions \citep{esposito24,li24}, which could occur locally at the jet--ISM interaction point. In this regard, the CO SLED of the integrated central region of ESO\,420-G13, detected with \textit{Herschel}/SPIRE \citep{kamenetzky16}, rises towards lower-J transitions, with a CO(7--6)/CO(12--11) flux ratio of 1.8, suggesting that extreme high-J excitation is not dominant on global scales.

Overall, these estimates support a scenario in which a substantial fraction ($\sim$32\%) of the outflowing molecular mass is invisible to conventional tracers such as CO(2--1). We favour the CO dissociation scenario, given the location of the blueshifted outflow at the edge of the coronal gas and X-ray emission, although extreme excitation cannot be fully excluded without high-angular resolution observations of mid- to high-J transitions. If CO is dissociated by jet-driven shocks or cosmic rays, strong emission from atomic or ionised carbon would be expected; future ALMA observations of the [\ion{C}{i}]$_{609}$ transition could provide key constraints on the properties of this CO-dark outflow.

\begin{figure}
\centering
\includegraphics[width = \columnwidth]{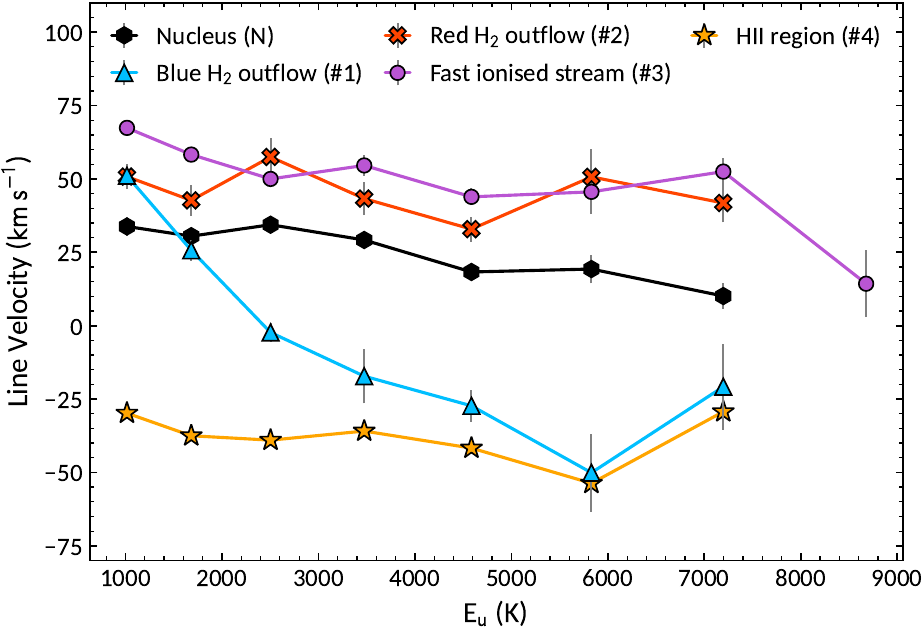}
\caption{Gaussian fit velocity vs upper state energy ($E_\text{u}$) for warm H$_2$ S(1) to S(8) rotational transitions extracted from different regions in ESO\,420-G13 (Fig\,\ref{subfig_hii}): active nucleus (black hexagons), warm H$_2$ blueshifted (blue triangles) and redshifted (red crosses) outflows, fast ionised gas stream (purple circles), and star-forming disc (orange stars).}\label{fig_EU_vel}
\end{figure}

A closer inspection of the warm H$_2$ kinematics in Figs.\,\ref{fig_H2spec} and \ref{fig_EU_vel} reveals that the blueshifted outflow behaves differently from the other regions, where the line velocities are broadly consistent. The blueshifted outflow exhibits a clear velocity stratification with increasing upper-level energy: while the H$_2$\,S(1) velocity ($\sim 50\,\kms$) is consistent with the rotational velocity of the cold molecular gas on the receding side of the disc (Fig.\,\ref{subfig_COH2_vel}), the line centroids shift by $\Delta v \sim -70\,\kms$ for S(4) to S(7). Although the MIRI/MRS spectral resolution prevents the detailed shock structure from being resolved, this gradient exceeds the uncertainties of the relative velocity calibration of a few $\kms$ (Table\,\ref{tab_h2kin}). Such excitation-dependent velocity offsets are predicted by shock models \citep[e.g.][]{villa-velez24} and have also been observed in protostellar H$_2$ jets, where shocks produce offsets of up to $\sim 20\,\kms$ \citep[e.g.][]{vleugels25}, suggesting that the blueshifted outflow is currently being impacted by a passing shock. This interpretation is further supported by its location at the edge of the extended coronal gas and X-ray emission, coinciding with the jet--ISM interaction zone. The absence of velocity stratification and the detection of CO(2--1) emission in the redshifted outflow suggest that this cloud may be in a cooling post-shock phase, in which CO molecules have already re-formed. Both outflows and the nucleus also show large observed velocity dispersions in most H$_2$ transitions ($\sigma_\mathrm{H2} \gtrsim 50\,\mathrm{\kms}$). Although overall no clear trend with $E_\mathrm{U}$ is seen, the blueshifted outflow shows a tentative increase in $\sigma_\mathrm{cor}$, suggesting that velocity stratification may be accompanied by increasing turbulence. The remaining regions are consistent with the instrumental resolution (Table\,\ref{tab_h2kin}).

To derive H$_2$ mass outflow rates we assumed a conical morphology with an opening solid angle $\Omega$ for the expanding wind, following \citet{cresci15}. The volume per unit solid angle in a sphere of radius $r$ is $r^3/3$, thus the total volume of a conical outflow is $V_\text{out} = \Omega r_\text{out}^3 / 3$. The mass outflow rate is given by $\dot{M}_\text{out} = n_\text{out} \, \Omega \, r_\text{out}^2 \, v_\text{out}$, where $n_\text{out} = M_\text{out} / V_\text{out}$ is the average gas density. Combining these expressions with the molecular gas masses derived previously, we obtain
\begin{equation}\label{eq_mdot}
    \dot{M}_\text{out} = \frac{M_\text{out}}{V_\text{out}} \, \Omega r_\text{out}^2 v_\text{out}
    = \frac{3 M_\text{out}}{\Omega r_\text{out}^3} \, \Omega r_\text{out}^2 v_\text{out}
    = 3 v_\text{out} \frac{M_\text{out}}{r_\text{out}}.
\end{equation}

The launching point of the outflow must be identified to determine $r_\text{out}$ and apply Eq.\,\ref{eq_mdot}. In our previous work, we associated the launching region with a forking point in the CO(2--1) emission, located $\sim 100\, \mathrm{pc}$ from the base of the redshifted outflow \citep{jafo20}. This interpretation is revised in light of the newly discovered blueshifted warm H$_2$ outflow and the expanding H$_2$ bubble, with the fast ionised gas stream located approximately at its centre ($\Delta\alpha \sim +1\farcs4$, $\Delta\delta \sim +0\farcs7$; Figs.\,\ref{subfig_s4_sig} and \ref{subfig_H2_sig}). We now consider two possible scenarios: \textit{i)} the outflows are launched from region~\#3 at the centre of the bubble; or \textit{ii)} they are directly launched from the nucleus. The choice of $r_\text{out}$ in each scenario leads to different values of the mass outflow rate, the momentum rate, and kinetic luminosity. The values obtained for these two scenarios are listed in Appendix\,\ref{app_outflow_props}, Table\,\ref{tab_winds}. Assuming a total warm plus cold H$_2$ mass of $\sim 4.0 \times 10^6\, \mathrm{M_\odot}$ and $8.7 \times 10^6\, \mathrm{M_\odot}$ for the outflows in regions~\#1 and \#2, respectively, we obtain outflow rates of $2.2$ and $11.2\, \mathrm{M_\odot\,yr^{-1}}$ if region~\#3 is the origin. If, instead, the outflows are launched from the nucleus, these rates and the derived quantities are reduced by factors of $1.7$ and $1.2$, respectively. The momentum rate carried by the outflowing molecular gas is given by $\dot{M}_\text{out}\, v_\text{out}$, where $v_\text{out}$ is measured from the velocity map in Fig.\,\ref{subfig_H2_vel}. This results in $2.8 \times 10^{33}\, \mathrm{erg\,cm^{-1}}$ and $11.3 \times 10^{33}\, \mathrm{erg\,cm^{-1}}$ for the H$_2$ outflows in regions~\#1 and \#2, respectively. Finally, the kinetic luminosity is obtained from
\begin{equation}
    L_\text{kin} = \frac{1}{2} \dot{M}_\text{out} \left(v^2_\text{out} + \kappa\,\sigma^2_\text{out}\right),
\end{equation}
where the intrinsic velocity dispersion $\sigma_\text{out}$ is multiplied by the geometric factor $\kappa$, which converts the observed 1D dispersion $\sigma_{\rm out}$ into a 3D mean-square velocity, $\langle v^2\rangle \sim \kappa\,\sigma_{\rm out}^2$. For isotropic turbulence, $\kappa = 3$ is commonly adopted \citep[e.g.][]{rodriguez-zaurin13}, while $\kappa = 2$ approximates anisotropic turbulence with two dominant degrees of freedom, as may be expected in a conical or cylindrical jet-driven flow. In regions with large $\sigma_\text{out}$, this turbulent term can substantially increase the inferred kinetic power. The resulting kinetic powers are $L_\text{kin} \sim 0.8 \times 10^{40}\, \mathrm{erg\,s^{-1}}$ for region~\#1 and $12.5 \times 10^{40}\, \mathrm{erg\,s^{-1}}$ for region~\#2 (see Section\,\ref{disc_coronal} for a comparison with the ionised gas).

\begin{figure*}
\centering
\subfigure[]{\includegraphics[width = \columnwidth]{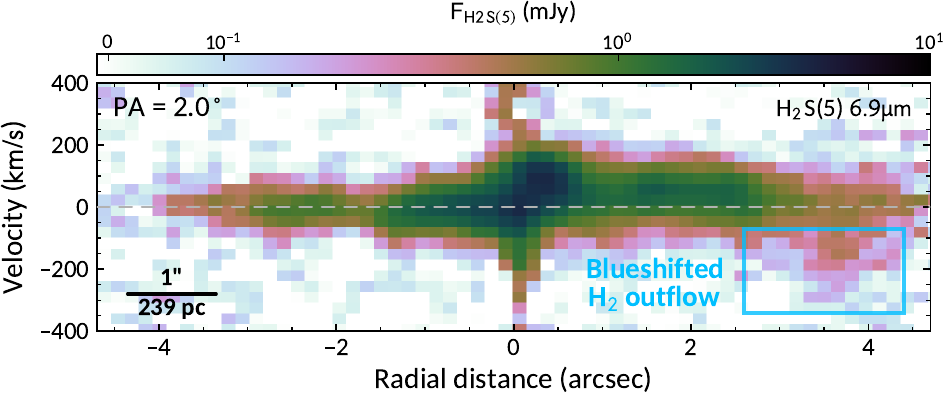}\label{subfig_pv2_h2}}~
\subfigure[]{\includegraphics[width = \columnwidth]{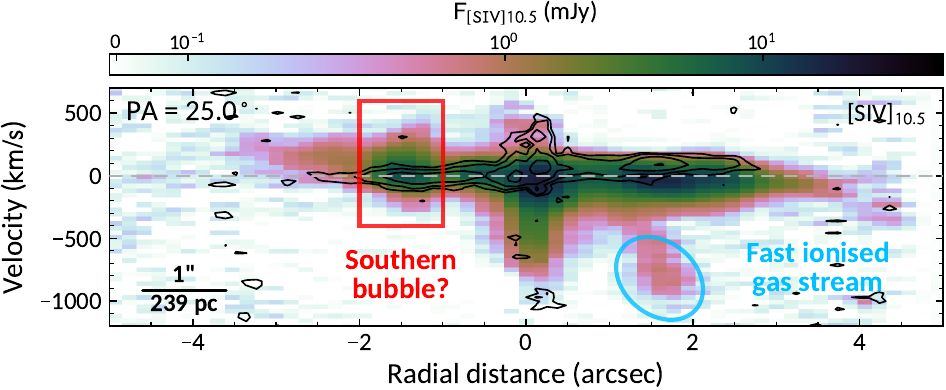}\label{subfig_pv25_s4}}\\[-0.1cm]
\subfigure[]{\includegraphics[width = \columnwidth]{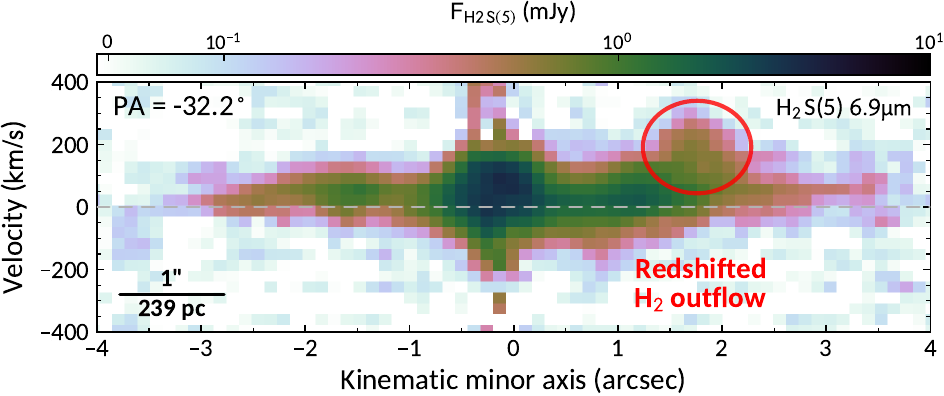}\label{subfig_pv32_h2}}~
\subfigure[]{\includegraphics[width = \columnwidth]{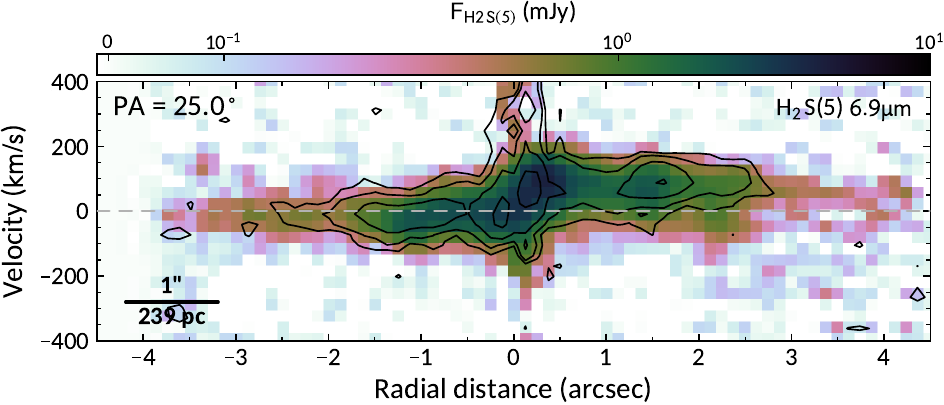}\label{subfig_pv25_h2}}\\[-0.2cm]
\caption{Position velocity diagrams for the H$_2$\,S(5) transition at $6.9\, \mathrm{\micron}$ \textbf{(a, c,} and \textbf{d)} and the [\ion{S}{iv}]$_{10.5}$ line \textbf{(b)}. The pseudo-slits are extracted from the line datacubes along the directions with $PA = 2^\circ$ and $-32\fdg2$ (kinematic minor axis), intersecting the blueshifted and redshifted warm molecular gas outflows, respectively, and along $PA = 25^\circ$, dissecting the coronal gas collimated emission. Pf$\alpha$ contours are shown in \textbf{(b)} and \textbf{(d)}. The molecular gas outflows appear as $\pm 250\, \mathrm{\kms}$ deviations from the disc rotation. The fast blueshifted ionised gas stream is clearly detected $\sim 1''$--$2''$ north of the nucleus, whereas enhanced turbulence appears at a similar distance in the southern direction, suggesting the presence of a coronal gas bubble at this location. The orientations of the three pseudo-slits are indicated in Fig.\,\ref{subfig_hii}.}\label{fig_pvmaps}
\end{figure*}

\subsection{Coronal wind mass, outflow rate, and energetics}\label{disc_coronal}
The absence of emission from low-ionisation species suggests over-ionisation in the collimated wind (see Pf$\alpha$ in Fig.\,\ref{subfig_pv25_s4}). Consequently, the lack of hydrogen recombination lines prevents us from estimating the ionised gas mass via case~B recombination, which relates the mass to the line intensity, electron density, and recombination rate \citep{osterbrock06}. Alternatively, the mass can be estimated from the intensity of collisionally excited transitions of the dominant ionic species of a given element (e.g. \citealt{ceci26} for the [\ion{Ne}{v}]$_{14.3}$ and [\ion{O}{iii}]$\lambda 5007$ lines), adopting a relative abundance with respect to hydrogen. The general methodology for this approach, assuming negligible collisional de-excitation, is described in Appendix\,\ref{app_ionmass}.

To estimate the ionised gas mass, we focus on the neon transitions detected in the MIRI/MRS range, which span multiple ionic species and therefore allow us to account for most of the neon mass. The detection of the collimated ionised wind in both the [\ion{Ne}{iii}]$_{15.6}$ and [\ion{Ne}{vi}]$_{7.7}$ lines is consistent with a relatively narrow temperature range, within which Ne$^{3+}$ and Ne$^{4+}$ are expected to be the dominant ionic species (Fig.\,\ref{fig_emiss}a). At $T_\text{e} \lesssim 20\,000\, \mathrm{K}$, the [\ion{Ne}{v}]$_{14.3,24.3}$ transitions dominate the cooling among the neon lines, because Ne$^{3+}$ lacks low-lying excited levels and therefore has no strong mid-IR transitions. In contrast, optical/UV transitions of Ne$^{3+}$ and Ne$^{4+}$ significantly contribute to cooling only at $\gtrsim 20\,000\, \mathrm{K}$, when the relevant energy levels become sufficiently populated (Fig.\,\ref{fig_emiss}b). The ionised gas mass can then be derived from the total [\ion{Ne}{v}]$_{14.3+24.3}$ flux ($F_{14.3+24.3}$), assuming a Ne$^{4+}$ ionic fraction of $f_\mathrm{Ne^{4+}} \sim 0.46$, a solar neon abundance ($X_\mathrm{Ne}=\log(\mathrm{Ne/H})=-3.94\,\mathrm{dex}$; \citealt{asplund21})
\begin{equation}\label{eq_mass_ne}
  M_\text{\sc Hii} = 8.06 \times 10^{16} \left( \frac{n_\text{e}}{\mathrm{cm^{-3}}} \right)^{-1} \left( \frac{D_\text{L}}{\mathrm{Mpc}} \right)^2 \left( \frac{F_{14.3+24.3}}{\mathrm{erg\,cm^{-2}\,s^{-1}}} \right) \, \mathrm{M_\odot}.
\end{equation}

For the integrated [\ion{Ne}{v}]$_{14.3+24.3}$ emission within the MIRI FoV, we derive a total ionised gas mass of $M^\text{wind}_\text{\sc Hii}$\,$\sim$\,$4.8 \times 10^5\, \mathrm{M_\odot}$ assuming $n_\text{e} \sim 100\, \mathrm{cm^{-3}}$, consistent with the low-density limit for the [\ion{Ne}{v}]$_{24.3}$/[\ion{Ne}{v}]$_{14.3}$ ratio (Fig.\,\ref{fig_dens}). This value does not include nuclear flux, which has a much higher density ($n_\text{e} \sim 2500\, \mathrm{cm^{-3}}$). For the fast ionised gas stream in region~\#3, we obtain $M^\text{stream}_\text{\sc Hii}$\,$\sim$\,$5.1 \times 10^3\, \mathrm{M_\odot}$ under the same assumption.

To test the validity of this approach, we applied the same method to the star-forming region~\#4. In this case, we assumed that the cooling is dominated by [\ion{Ne}{ii}]$_{12.8}$, the only IR transition of the most abundant neon ionic species in star-forming regions with low ionisation parameters \citep{ho07}. In contrast to the coronal gas regions, Pf$\alpha$ is detected in the spectrum of region~\#4 (see Fig.\,\ref{fig_discspec}), allowing a direct comparison between the ionised gas mass derived from collisional line emission and that obtained using the classical recombination-line method of \citet{osterbrock06}. For region~\#4, we obtain $M_\text{\sc Hii} \sim 1.1 \times 10^6\, \mathrm{M_\odot}$ from Pf$\alpha$ and $\sim 2.1 \times 10^6\, \mathrm{M_\odot}$ from [\ion{Ne}{ii}]$_{12.8}$. This comparison suggests that the ionised gas masses derived from collisionally excited lines using this method are reliable to within a factor of a few. This agreement supports the reliability of the neon-based approach for estimating ionised gas masses in regions where recombination lines are undetected, as in the case of the collimated coronal wind in ESO\,420-G13.

Building on the ionised gas mass estimates derived above, we now assess the outflow energetics for the extended ionised wind and the fast ionised gas stream. The dynamical time ($t_\text{d}$) of the extended wind can be estimated by assuming a radius of $r_\text{out} \sim 870\, \mathrm{pc}$ and an expansion velocity of $v_\text{out} \gtrsim 80\, \mathrm{\kms}$, which yields $t_\text{d} = r_\text{out} / v_\text{out} \lesssim 11\, \mathrm{Myr}$. This timescale is consistent with the duration of a typical AGN outburst \citep[$10^4$--$10^7\, \mathrm{yr}$;][]{husemann22}, suggesting that the extended collimated emission in ESO\,420-G13 may have been produced during the most recent episode of nuclear activity. Since the observed line-of-sight velocity provides only a lower limit to the true outflow velocity, the dynamical timescales should be regarded as upper limits, whereas the mass outflow rates are lower limits.

Using Eq.\,\ref{eq_mdot}, and adopting $M_\text{\sc Hii} \sim 4.8 \times 10^5\, \mathrm{M_\odot}$ and $r_\text{out} \sim 870\, \mathrm{pc}$ for the extended collimated wind, we derive a mass outflow rate of $\dot{M}^\text{wind}_\text{out} \sim 0.14\, \mathrm{M_\odot\,yr^{-1}}$. For the fast ionised gas stream, with $M^\text{stream}_\text{\sc Hii} \sim 5.1 \times 10^3\, \mathrm{M_\odot}$ and $r_\text{out} \sim 210\, \mathrm{pc}$ (based on its spatial extent; Fig.\,\ref{subfig_s4_sig}), we obtain $\dot{M}^\text{stream}_\text{out} \sim 0.06\, \mathrm{M_\odot\,yr^{-1}}$. These ionised gas outflow rates and their associated momentum rates are well below those for warm molecular gas (Table\,\ref{tab_winds}), reflecting the smaller gas masses in the ionised phase. However, the high velocity and large velocity dispersion of the fast ionised stream imply a kinetic power of $L_\text{kin} \sim 1.1 \times 10^{40}\, \mathrm{erg\,s^{-1}}$, comparable to that of the molecular gas outflows. This supports a scenario in which an expanding H$_2$ bubble is driven by the fast ionised stream originating in region~\#3. In contrast, an alternative scenario in which molecular outflows are launched from the nucleus would yield significantly lower kinetic power for the extended ionised gas (Table\,\ref{tab_winds}).

The coronal gas does not display high-velocity streams south of the nucleus, although a region of enhanced velocity dispersion ($\sigma \sim 100\, \mathrm{\kms}$) is found at $\sim -1\farcs5$ ($\sim 360\, \mathrm{pc}$), as shown by the position-velocity (PV) map of the [\ion{S}{iv}]$_{10.5}$ emission in Fig.\,\ref{subfig_pv25_s4}. This feature likely traces an ionised gas bubble that may eventually break out and drive molecular outflows similar to those observed to the north.

\subsection{Powerful kinetic feedback from a faint nucleus}\label{disc_feedback}
The jet in ESO\,420-G13 has an estimated kinetic power of $\sim$\,$4 \times 10^{42}\,\mathrm{erg\,s^{-1}}$ \citep{jafo20}, derived from the \citet{heckman14} correlation between the monochromatic $1.4\,\mathrm{GHz}$ luminosity and the work required to inflate X-ray cavities in radio galaxies. The lack of extended radio emission in MeerKAT $1.4\,\mathrm{GHz}$ observations, which show only a compact nucleus ($<6''$; \citealt{condon21}), suggests that the jet is still in an early stage of development. This is consistent with the detection of ionised gas with kinematic distortions, a feature more commonly found in young radio AGNs \citep{kukreti24}.

The mechanical power available from the jet contrasts with the comparatively weak radiative output of the nucleus, with $L_\mathrm{2-10\,keV} \sim 10^{40}\,\mathrm{erg\,s^{-1}}$ \citep{lehmer10} and a bolometric luminosity of $\sim 3 \times 10^{43}\,\mathrm{erg\,s^{-1}}$, estimated from the nuclear luminosities of [\ion{Ne}{v}]$_{14.3}$ and [\ion{O}{iv}]$_{25.9}$ (Table\,\ref{tab_ionflux}) using the calibration from \citet{spinoglio24}. Although a radiatively driven wind cannot be entirely ruled out, it does not naturally explain the highly collimated morphology and pronounced bend of the coronal emission, the launching of the fast ionised gas stream $\sim 370\,\mathrm{pc}$ from the nucleus, or the presence of a seemingly undisturbed reservoir of cold molecular gas rotating close to the central black hole. This picture is reinforced by mid-IR excitation diagnostics. The faintness of the nucleus is also reflected in the mid-IR line ratios measured within the $0\farcs7$ nuclear aperture (Table\,B.2). In particular, [\ion{Ne}{iii}]$_{15.6}$/[\ion{Ne}{ii}]$_{12.8}\sim 0.26$ is only marginally higher than the value obtained from the MIRI FoV-integrated spectrum ($\sim 0.23$). Similarly, the nuclear [\ion{Ar}{iii}]$_{9.0}$/[\ion{Ar}{ii}]$_{7.0}\sim 0.17$ is low compared to typical Seyfert nuclei. Together, these ratios are more consistent with low-luminosity AGN excitation (\citealt{goold26}; Acharya et al., in prep.), shock excitation \citep{ceci26}, or cosmic ray excitation \citep{koutsoumpou25b} than with photoionisation from a bright nucleus.

By comparing the total molecular plus ionised outflow kinetic power ($\sim$$1.4 \times 10^{41}\,\mathrm{erg\,s^{-1}}$) with the jet power estimated above, we infer a jet--ISM coupling efficiency of $\sim$3.8\%, consistent with observational estimates \citep[0.1--5\%;][]{harrison18,fluetsch19}. Higher theoretical values, such as $10$--$40$\% in hydrodynamical simulations \citep{wagner12} or $5$--$20$\% required in cosmological models to reproduce the $M_\mathrm{BH}$--$\sigma_*$ relation \citep{dimatteo05,schaye15,weinberger17}, should be regarded as upper limits, since only a fraction of the injected AGN energy is converted into outflow kinetic power.

The interaction between the jet and the ISM in ESO\,420-G13 has already removed $\sim$5\% of the central molecular gas reservoir via outflows in regions~\#1 and~\#2. To assess the impact on ongoing star formation, we estimate the mass-loading factor $\eta = \dot{M}_\mathrm{out}/\mathrm{SFR}$. The star-formation rate (SFR) derived from the integrated [\ion{Ne}{ii}]$_{12.8}$ emission \citep{zhuang19,mordini21}, adopting a fixed [\ion{Ne}{iii}]$_{15.6}$/[\ion{Ne}{ii}]$_{12.8}=0.05$ from region~\#4 to minimise jet contamination, yields $\mathrm{SFR} \sim 15$--$25\,\mathrm{M_\odot\,yr^{-1}}$. An independent estimate from the integrated PAH\,$11.3\,\mu$m flux, $(8.94 \pm 1.0) \times 10^{-12}\, \mathrm{erg\,s^{-1}\,cm^{-2}}$, measured by fitting asymmetric Drude profiles on a polynomial continuum, gives $\mathrm{SFR} = 4.1 \pm 0.7\,\mathrm{M_\odot\,yr^{-1}}$ \citep{shipley16}. The discrepancy likely reflects PAH suppression by the AGN and/or residual contamination of [\ion{Ne}{ii}] by jet-excited gas. Taking these estimates as lower and upper bounds gives $\eta \sim 0.5$--$3.2$, with $\eta \gtrsim 0.9$ for two of them. This suggests that the outflow rate may exceed the SFR and that feedback could efficiently deplete the remaining gas in the central region of ESO\,420-G13.


\section{Summary}\label{summary}

We present JWST/MIRI mid-IR integral-field spectroscopy of the post-starburst galaxy ESO\,420-G13, combined with ALMA CO(2--1) observations, to investigate how AGN-driven mechanical energy couples to the multiphase ISM on sub-kiloparsec scales. The spatially resolved MIRI diagnostics map the morphology, excitation, and kinematics of warm H$_2$ molecular gas and highly ionised coronal gas, enabling a direct comparison with the cold molecular component traced by CO.

The jet trail is exposed by collimated coronal-line emission and extended X-ray emission, reaching $\gtrsim 870\,\mathrm{pc}$ from the nucleus. The jet--ISM interaction is strongest $\sim 370\,\mathrm{pc}$ to the north, where a fast ionised gas stream emerges perpendicular to the jet axis and coincides with a pronounced bend in the jet trajectory. Surrounding this stream, the warm H$_2$ shows enhanced velocity dispersion ($\sigma_{\mathrm{H_2S(5)}} \gtrsim 80\,\mathrm{km\,s^{-1}}$), consistent with an expanding molecular bubble. Two warm-H$_2$ outflows are detected at the edges of this structure (regions~\#1 and \#2). The blueshifted outflow is CO-dark and shows velocity stratification across the H$_2$ rotational ladder, pointing to ongoing shocks and/or cosmic-ray heating that can dissociate CO and strongly modify the CO-to-H$_2$ conversion. Enhanced coronal turbulence is also detected along the southern jet at $\sim 240\,\mathrm{pc}$, hinting at a less developed analogue of the northern structure.

We estimate a combined molecular plus ionised outflow kinetic power of $\sim 1.4 \times 10^{41}\,\mathrm{erg\,s^{-1}}$. Compared to the jet kinetic power inferred from radio luminosity ($\sim 4 \times 10^{42}\,\mathrm{erg\,s^{-1}}$), this implies a jet--ISM coupling efficiency of $\sim 3.8\%$. This level of coupling is remarkable given the modest nuclear luminosity, and it illustrates that radiatively faint nuclei can drive substantial feedback to their host galaxies.

The interpretation also aligns with the post-starburst nature of ESO\,420-G13, where the final nuclear star-formation episodes may coincide with the onset of AGN feedback. Most of the cold molecular gas within the central kiloparsec ($\sim 3 \times 10^8\,\mathrm{M_\odot}$) remains in ordered rotation with low turbulence, yet $\sim 5\%$ of the central molecular reservoir has already been expelled, which implies a mass-loading factor of $\eta \gtrsim 1$. Nevertheless, the warm H$_2$ in the disc is turbulent and includes high-temperature components ($T_\mathrm{ex}\gtrsim 1700\,\mathrm{K}$), conditions unfavourable for future star formation, suggesting that jet-driven mechanical feedback is beginning to affect the central ISM following the starburst phase.

This work highlights the unique power of mid-IR imaging spectroscopy for diagnosing AGN jet feedback in galaxy nuclei. Without JWST/MIRI's spatially resolved coronal and H$_2$ diagnostics, the jet-driven coupling and much of the kinetic energy budget in ESO\,420-G13, and therefore its feedback impact, would likely remain hidden to conventional tracers.


\begin{acknowledgements}
The authors dedicate this work to the memory of Ángel Rolland Quintanilla and Francisco Sánchez Martínez. Without people like them, who paved the way for later generations, careers such as ours would simply not have been possible. The authors are grateful to the anonymous referee, whose thorough and constructive comments significantly improved the quality of this paper. JAFO, MPS, AHC, EPM, JMV, BPD, and RA acknowledge financial support from the Spanish Ministry of Science, Innovation and Universities (MICIU/AEI/10.13039/501100011033). This work was supported by grants PID2021-124918NB-C44 (JAFO and AHC), PID2023-147386NB-I00 (JAFO, AHC, and RA), and PID2023-146667NB-I00 (MPS), and PID2022-136598NB-C32 (EPM, JMV, and BPD), funded by MICIU/AEI and ERDF/EU, and by grant RYC2021-033094-I (MPS). JAFO, RA, and MPS also acknowledge support from MICIU/AEI and the European Union NextGenerationEU/PRTR through grants CNS2023-145339 (JAFO and RA) and CNS2023-145506 (MPS). JAFO further acknowledges support from MICIU and the European Union -- NextGenerationEU through the Recovery and Resilience Facility project ICTS-MRR-2021-03-CEFCA. EH has received funding from the European Union's Horizon Europe research and innovation programme under grant agreement No. 101188037 (AtLAST2). This research has made use of data obtained from the \textit{Chandra} Data Archive provided by the \textit{Chandra} X-ray Center (CXC). This work made use of Astropy\footnote{\url{http://www.astropy.org}}: a community-developed core Python package and an ecosystem of tools and resources for astronomy \citep{astropy22}.
\end{acknowledgements}

\bibliographystyle{aa}
\bibliography{wild}

\begin{appendix}

\onecolumn
\nolinenumbers

\section{Extracted spectra for selected regions}\label{app_specs}

\begin{figure*}[ht!!!]
\centering
\includegraphics[width = 0.9\textwidth]{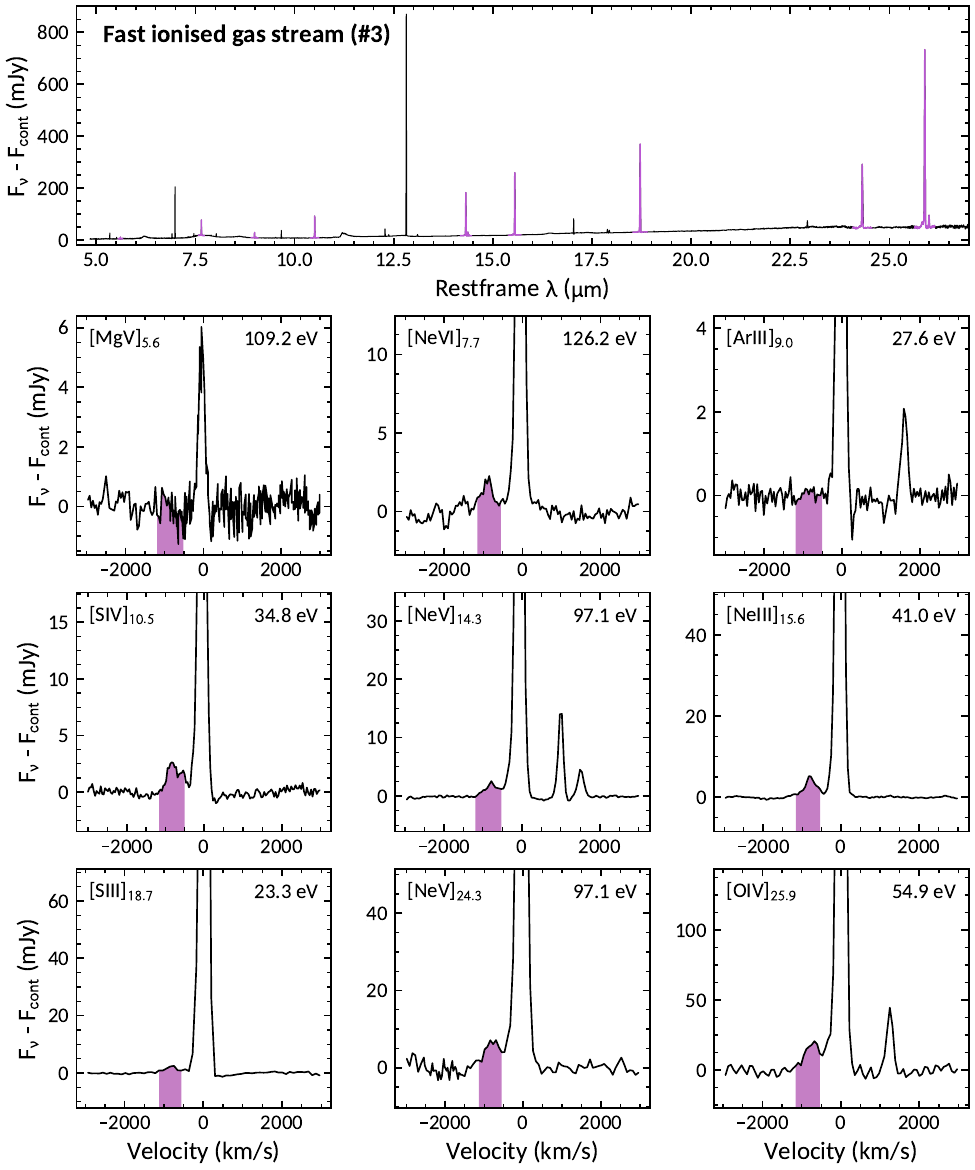}
\caption{The top panel shows the continuum subtracted spectrum (aperture radius of $0\farcs7$, background estimated at radii $0\farcs8$--$1\farcs2$) at the position of the fast ionised gas stream \#3, as indicated in Fig.\,\ref{fig_regions}. The lower panels show several transitions where the broad blueshifted component ($-1200$ to $-500\, \mathrm{\kms}$) is detected (purple-shaded area), after continuum subtraction. No broad component is detected for the [\ion{Mg}{v}]$_{5.6}$ or the [\ion{Ar}{iii}]$_{ 9.0}$ lines, although a marginal detection is seen in [\ion{S}{iii}]$_{18.7}$. Note the [\ion{Na}{iv}]$_{9.0}$ line detected next to [\ion{Ar}{iii}]$_{9.0}$, the [\ion{Cl}{ii}]$_{14.3}$ and [\ion{Na}{iv}]$_{14.4}$ lines next to [\ion{Ne}{v}]$_{14.3}$, and the [\ion{Fe}{ii}]$_{26}$ line next to [\ion{O}{iv}]$_{25.9}$.}\label{fig_windspec}
\end{figure*}

\begin{figure*}[ht!!!]
\centering
\includegraphics[width = 0.9\textwidth]{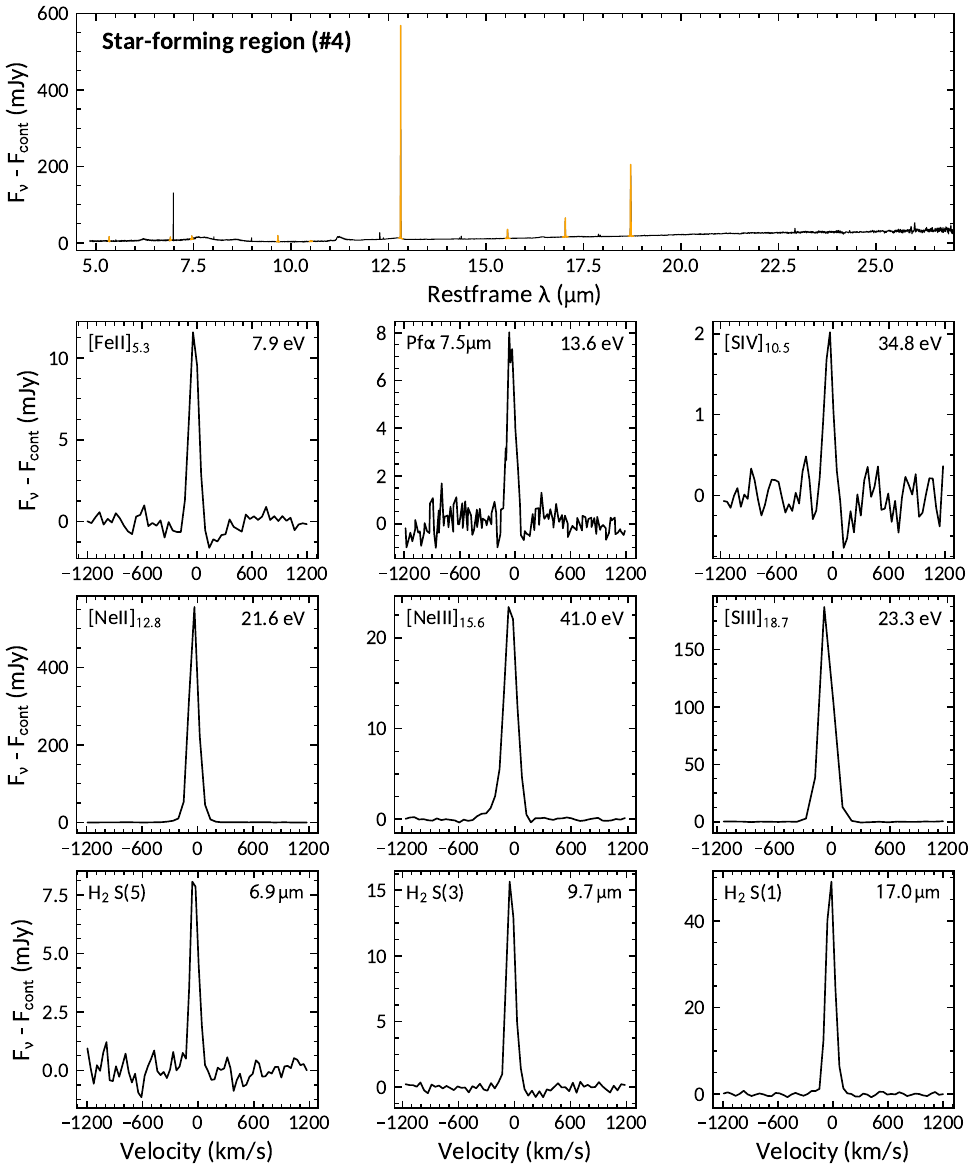}
\caption{The top panel shows the continuum-subtracted spectrum (aperture radius of $0\farcs7$, background estimated at radii $0\farcs8$--$1\farcs2$) at the position \#4, corresponding to a star-forming region in the disc of ESO\,420-G13 without overlap with the coronal gas emission (Fig.\,\ref{fig_regions}). The lower panels show typical low-ionisation and warm H$_2$ rotational transitions in the mid-IR range, after continuum subtraction.}\label{fig_discspec}
\end{figure*}

\FloatBarrier
\twocolumn


\onecolumn
\section{Line fluxes}\label{app_fluxes}
This appendix compiles the emission-line fluxes measured from the continuum-subtracted \textit{JWST}/MIRI-MRS spectra of ESO\,420-G13. Line fluxes were measured using custom \texttt{python}\footnote{\url{https://www.python.org}} routines by integrating the line profile after subtracting the continuum level, which was determined from a one-dimensional polynomial fit to the flux density distribution of the adjacent continuum on both sides of the line. The values are provided for the nuclear aperture, the selected regions defined in Fig.\,\ref{subfig_hii}, and the integrated MIRI field of view, and include both the pure rotational H$_2$ transitions (Table\,\ref{tab_h2flux}) and the main ionic fine-structure lines detected across the mid-IR range (Table\,\ref{tab_ionflux}). Upper limits correspond to a $3\sigma$ confidence level. These measurements form the basis of the analysis presented in Sections\,\ref{results} and \ref{discuss}, including the rotational diagrams, excitation diagnostics, and estimates of the ionised and molecular gas properties.


\begin{table*}[h!!!]
\caption{H$_2$ rotational emission-line fluxes in ESO\,420-G13.}\label{tab_h2flux}
\small
\centering
\begin{tabular}{lcccccccc}
 \bf Line & \bf Wavelength  & \bf $E_\text{u}$ & \bf Nucleus & \bf Blueshifted   & \bf Redshifted    & \bf Fast ionised & \bf Star-forming & \bf MIRI/MRS \\
          &                 &                  & \bf (N)     & \bf outflow (\#1) & \bf outflow (\#2) & \bf stream (\#3) & \bf region (\#4) & \bf FoV      \\
          & [$\mathrm{\micron}$] & [K] & \multicolumn{6}{c}{[$\times 10^{-14}\, \mathrm{erg\,s^{-1}\,cm^{-2}}$]} \\
  \hline \\[-0.2cm]
  H$_2$\,S(1) & 17.034845756 & 1015.1 &  $0.59 \pm 0.04$ & $0.22 \pm 0.01$ & $0.55 \pm 0.02$ & $0.33 \pm 0.04$ & $0.35 \pm 0.01$ & $14.54 \pm 0.07$ \\
  H$_2$\,S(2) & 12.278611991 & 1681.6 &  $0.48 \pm 0.03$ & $0.11 \pm 0.01$ & $0.32 \pm 0.01$ & $0.32 \pm 0.02$ & $0.20 \pm 0.01$ &  $6.91 \pm 0.09$ \\
  H$_2$\,S(3) &  9.664910918 & 2503.7 &  $0.74 \pm 0.02$ & $0.20 \pm 0.01$ & $0.44 \pm 0.01$ & $0.35 \pm 0.01$ & $0.17 \pm 0.01$ &  $9.68 \pm 0.07$ \\
  H$_2$\,S(4) &  8.025041036 & 3474.5 &  $0.42 \pm 0.02$ & $0.07 \pm 0.01$ & $0.16 \pm 0.01$ & $0.16 \pm 0.02$ & $0.09 \pm 0.01$ &  $3.60 \pm 0.05$ \\
  H$_2$\,S(5) &  6.909508549 & 4586.1 &  $0.77 \pm 0.03$ & $0.16 \pm 0.01$ & $0.25 \pm 0.02$ & $0.24 \pm 0.02$ & $0.12 \pm 0.02$ &  $6.0 \pm 0.2$ \\
  H$_2$\,S(6) &  6.108563840 & 5829.8 &  $0.16 \pm 0.01$ & $0.07 \pm 0.01$ & $0.03 \pm 0.01$ & $0.06 \pm 0.01$ & $0.02 \pm 0.01$ &  $1.48 \pm 0.06$ \\
  H$_2$\,S(7) &  5.511183259 & 7196.7 &  $0.48 \pm 0.03$ & $0.04 \pm 0.02$ & $0.10 \pm 0.02$ & $0.09 \pm 0.01$ & $0.05 \pm 0.02$ &  $2.8 \pm 0.1$ \\
  H$_2$\,S(8) &  5.053115155 & 8677.1 &          $<0.09$ &        $< 0.06$ &        $< 0.07$ & $0.06 \pm 0.02$ &        $< 0.06$ &         $< 0.5$ \\
  \hline
\end{tabular}
\tablefoot{Line rest-frame wavelengths and upper level excitation temperatures from \citet{roueff19}.}
\end{table*}

\begin{table*}[h!]
\caption{H$_2$ rotational emission-line kinematic properties in ESO\,420-G13.}\label{tab_h2kin}
\scriptsize
\setlength{\tabcolsep}{3.pt}
\centering
\begin{tabular}{l *{15}{S[table-format=-2.0(2)]}}
 \textbf{Line} &
 \multicolumn{3}{c}{\textbf{Nucleus (N)}} &
 \multicolumn{3}{c}{\textbf{Blueshifted outflow (\#1)}} &
 \multicolumn{3}{c}{\textbf{Redshifted outflow (\#2)}} &
 \multicolumn{3}{c}{\textbf{Fast ionised stream (\#3)}} &
 \multicolumn{3}{c}{\textbf{Star-forming region (\#4)}} \\
 &
 \multicolumn{1}{c}{\textbf{$v$}} &
 \multicolumn{1}{c}{\textbf{$\sigma_\mathrm{obs}$}} &
 \multicolumn{1}{c}{\textbf{$\sigma_\mathrm{cor}$}} &
 \multicolumn{1}{c}{\textbf{$v$}} &
 \multicolumn{1}{c}{\textbf{$\sigma_\mathrm{obs}$}} &
 \multicolumn{1}{c}{\textbf{$\sigma_\mathrm{cor}$}} &
 \multicolumn{1}{c}{\textbf{$v$}} &
 \multicolumn{1}{c}{\textbf{$\sigma_\mathrm{obs}$}} &
 \multicolumn{1}{c}{\textbf{$\sigma_\mathrm{cor}$}} &
 \multicolumn{1}{c}{\textbf{$v$}} &
 \multicolumn{1}{c}{\textbf{$\sigma_\mathrm{obs}$}} &
 \multicolumn{1}{c}{\textbf{$\sigma_\mathrm{cor}$}} &
 \multicolumn{1}{c}{\textbf{$v$}} &
 \multicolumn{1}{c}{\textbf{$\sigma_\mathrm{obs}$}} &
 \multicolumn{1}{c}{\textbf{$\sigma_\mathrm{cor}$}} \\
 & \multicolumn{3}{c}{[$\mathrm{\kms}$]} &
 \multicolumn{3}{c}{[$\mathrm{\kms}$]} &
 \multicolumn{3}{c}{[$\mathrm{\kms}$]} &
 \multicolumn{3}{c}{[$\mathrm{\kms}$]} &
 \multicolumn{3}{c}{[$\mathrm{\kms}$]} \\
  \hline \\[-0.2cm]
  H$_2$\,S(1) &  34(3) & 74(3) & 52(5) & 51(2) & 56(2) & 19(9) & 51(4) & 71(4) & 47(7) & 67(2) & 46(2) & \multicolumn{1}{c}{--} & -30(1) & 46(1) & \multicolumn{1}{c}{--} \\
  H$_2$\,S(2) &  31(2) & 76(2) & 63(3) & 26(4) & 64(4) & 48(5) & 43(5) & 75(5) & 62(7) & 58(1) & 54(1) & 34(3) & -38(1) & 49(1) & 24(4) \\
  H$_2$\,S(3) &  34(2) & 75(2) & 65(3) & -2(3) & 67(3) & 55(4) & 58(6) & 84(6) & 74(7) & 50(2) & 50(2) & 33(4) & -39(1) & 44(1) & 21(4) \\
  H$_2$\,S(4) &  29(3) & 73(3) & 64(4) & -17(9) & 76(9) & 70(10) & 43(6) & 68(6) & 58(7) & 55(4) & 51(4) & 36(5) & -36(1) & 43(1) & 24(4) \\
  H$_2$\,S(5) &  18(2) & 78(2) & 70(2) & -27(6) & 80(6) & 72(6) & 33(4) & 61(4) & 51(5) & 44(3) & 51(3) & 38(4) & -42(2) & 42(2) & 24(4) \\
  H$_2$\,S(6) &  19(5) & 71(5) & 62(6) & -50(10) & 80(10) & 70(20) & 50(10) & 50(10) & 30(10) & 46(8) & 55(8) & 40(10) & -54(7) & 33(7) & \multicolumn{1}{c}{--} \\
  H$_2$\,S(7) &  10(4) & 80(4) & 73(5) & -21(10) & 70(20) & 70(20) & 42(7) & 52(7) & 41(9) & 53(5) & 46(5) & 32(7) & -29(6) & 41(6) & 30(10) \\
  H$_2$\,S(8) & \multicolumn{1}{c}{--} & \multicolumn{1}{c}{--} & \multicolumn{1}{c}{--} &
                  \multicolumn{1}{c}{--} & \multicolumn{1}{c}{--} & \multicolumn{1}{c}{--} &
                  \multicolumn{1}{c}{--} & \multicolumn{1}{c}{--} & \multicolumn{1}{c}{--} &
                  10(10) & 90(10) & 80(10) &
                  \multicolumn{1}{c}{--} & \multicolumn{1}{c}{--} & \multicolumn{1}{c}{--} \\
  \hline
\end{tabular}
\tablefoot{Velocity ($v$), observed velocity dispersion ($\sigma_\mathrm{obs}$), and intrinsic velocity dispersion ($\sigma_\mathrm{cor}$) of the H$2$ rotational emission-line profiles in ESO\,420-G13. The values of $\sigma_\mathrm{cor}$ are obtained by correcting $\sigma_\mathrm{obs}$ in quadrature for the MIRI/MRS instrumental resolution, calculated using eq.\,1 of \citet{argyriou23}. A conservative fractional uncertainty of 5\% is adopted for the instrumental resolution.}
\end{table*}

\begin{landscape}
\begin{table*}[h!!!]
\caption{Ionic line fluxes.}\label{tab_ionflux}
\small
\centering
\begin{tabular}{lccccccccc}
 \bf Line & \bf Wavelength & \bf IP & \bf $n_\text{crit}$ & \bf Nucleus & \bf Blueshifted   & \bf Redshifted    & \bf Fast ionised & \bf Star-forming & \bf MIRI/MRS \\
          &                &        &                     & \bf (N)     & \bf outflow (\#1) & \bf outflow (\#2) & \bf stream (\#3) & \bf region (\#4) & \bf FoV      \\

          & [$\mathrm{\micron}$] & [eV] & [$\mathrm{cm^{-3}}$] & \multicolumn{6}{c}{[$\times 10^{-14}\, \mathrm{erg\,s^{-1}\,cm^{-2}}$]} \\
\hline \\[-0.2cm]
  {[\ion{Fe}{ii}]}   &  5.340169 &   7.90 & $3.007 \times 10^2$ &  $1.01 \pm 0.02$ &        $< 0.06$ &  $0.17 \pm 0.02$ &  $0.56 \pm 0.02$ & $0.20 \pm 0.02$ &  $9.52 \pm 0.15$ \\
  {[\ion{Fe}{viii}]} &    5.4466 & 124.98 & $2.553 \times 10^6$ &  $0.39 \pm 0.05$ &        $< 0.09$ &         $< 0.07$ & $0.054 \pm 0.014$ &  $< 0.09$ &  $0.92 \pm 0.22$ \\
  {[\ion{Mg}{vii}]}  &    5.5032 & 186.76 & $3.377 \times 10^6$ &  $0.32 \pm 0.03$ &        $< 0.06$ &         $< 0.06$ &  $0.03 \pm 0.01$ &        $< 0.06$ &  $0.64 \pm 0.12$ \\
  {[\ion{Mg}{v}]}    &   5.60985 & 109.27 & $4.011 \times 10^6$ &  $0.70 \pm 0.04$ &        $< 0.06$ &         $< 0.06$ &  $0.17 \pm 0.02$ &        $< 0.09$ &  $2.10 \pm 0.21$ \\
  {[\ion{Ni}{ii}]}   &     6.636 &   7.64 & $1.316 \times 10^6$ &  $0.31 \pm 0.04$ &        $< 0.06$ &         $< 0.06$ & $0.037 \pm 0.014$ &  $< 0.06$ &  $1.30 \pm 0.12$ \\
  {[\ion{Ar}{ii}]}   &  6.985274 &  15.76 & $4.169 \times 10^5$ &  $7.38 \pm 0.04$ & $0.12 \pm 0.02$ &  $1.04 \pm 0.02$ &  $3.64 \pm 0.03$ & $1.99 \pm 0.02$ & $49.69 \pm 0.14$ \\
  {[\ion{Na}{iii}]}  &    7.3177 &  47.29 & $6.397 \times 10^6$ &  $0.24 \pm 0.04$ &        $< 0.03$ &         $< 0.06$ & $0.12 \pm 0.03$ &  $< 0.06$ &  $0.43 \pm 0.13$ \\
  Pf\,$\alpha$       & 7.4598577 &  13.60 &                  -- &  $0.34 \pm 0.02$ &        $< 0.03$ & $0.05 \pm 0.01$ &   $0.15 \pm 0.02$ & $0.11 \pm 0.01$ &  $2.61 \pm 0.11$ \\
  Hu\,$\beta$        & 7.5024932 &  13.60 &                  -- &         $< 0.06$ &        $< 0.03$ &         $< 0.06$ & $0.05 \pm 0.02$ & $0.06 \pm 0.02$ & $0.85 \pm 0.14$ \\
  {[\ion{Ne}{vi}]}   &    7.6524 & 126.21 & $6.302 \times 10^5$ &  $5.44 \pm 0.08$ & $0.32 \pm 0.02$ & $0.08 \pm 0.04$ & $1.68 \pm 0.09$ & $< 0.09$ & $14.06 \pm 0.24$ \\
  {[\ion{Ne}{vi}]$_\text{b}$} & 7.6524 & 126.21 & $6.302 \times 10^5$ &         -- &              -- &               -- &  $0.14 \pm 0.05$ &              -- &               -- \\
  {[\ion{Ar}{v}]}    &    7.9016 &  59.81 & $1.596 \times 10^5$ &  $0.10 \pm 0.02$ &        $< 0.03$ &         $< 0.06$ & $0.10 \pm 0.03$ &  $< 0.03$ &         $< 0.42$ \\
  {[\ion{Ar}{iii}]}  &   8.99138 &  27.63 & $1.890 \times 10^5$ &  $1.27 \pm 0.03$ & $0.08 \pm 0.01$ &  $0.04 \pm 0.01$ &  $0.49 \pm 0.02$ & $0.11 \pm 0.01$ &  $4.54 \pm 0.10$ \\
  {[\ion{Ar}{iii}]$_\text{b}$} & 8.99138 & 27.63 & $1.890 \times 10^5$ &        -- &              -- &               -- &         $< 0.03$ &              -- &               -- \\
  {[\ion{Na}{iv}]}   &     9.041 &  71.62 & $9.976 \times 10^5$ &         $< 0.06$ &        $< 0.03$ &         $< 0.03$ &  $0.05 \pm 0.01$ &        $< 0.03$ &         $< 0.21$ \\
  {[\ion{Fe}{vii}]}  &    9.5267 &  99.10 & $5.529 \times 10^5$ &  $0.12 \pm 0.01$ &        $< 0.03$ &         $< 0.03$ &  $0.06 \pm 0.01$ &        $< 0.03$ &         $< 0.18$ \\
  {[\ion{S}{iv}]}    &   10.5105 &  34.79 & $5.596 \times 10^4$ &  $2.34 \pm 0.03$ & $0.32 \pm 0.01$ &         $< 0.03$ &  $1.55 \pm 0.02$ & $< 0.015 \pm 0.005$ &  $8.93 \pm 0.10$ \\
  {[\ion{S}{iv}]$_\text{b}$} & 10.5105 & 34.79 & $5.596 \times 10^4$ &          -- &              -- &               -- &  $0.11 \pm 0.01$ &              -- &               -- \\
  Hu\,$\alpha$       & 12.371898 &  13.60 &                  -- &  $0.10 \pm 0.02$ &        $< 0.03$ &         $< 0.03$ &  $0.07 \pm 0.02$ & $0.04 \pm 0.01$ &  $0.99 \pm 0.05$ \\
  {[\ion{Ne}{ii}]}   &  12.81355 &  21.56 & $6.289 \times 10^5$ & $20.32 \pm 0.06$ & $0.68 \pm 0.01$ &  $3.31 \pm 0.03$ & $10.00 \pm 0.06$ & $5.54 \pm 0.02$ &  $134.4 \pm 0.3$ \\
  {[\ion{Ar}{v}]}    &   13.1022 &  59.81 & $2.918 \times 10^4$ &  $0.03 \pm 0.02$ & $0.020 \pm 0.003$ &  $< 0.03$ &  $0.11 \pm 0.02$ &        $< 0.03$ &  $0.44 \pm 0.04$ \\
  {[\ion{Ne}{v}]}    &   14.3217 &  97.12 & $3.238 \times 10^4$ &  $3.50 \pm 0.05$ & $0.45 \pm 0.01$ &         $< 0.06$ &  $2.33 \pm 0.03$ & $0.029 \pm 0.005$ & $14.24 \pm 0.06$ \\
  {[\ion{Ne}{v}]$_\text{b}$} & 14.3217 &  97.12 & $3.238 \times 10^4$ &         -- &              -- &               -- &  $0.07 \pm 0.02$ &              -- &               -- \\
  {[\ion{Cl}{ii}]}   &   14.3678 &  12.97 & $3.896 \times 10^4$ &  $0.27 \pm 0.02$ & $0.02 \pm 0.01$ &  $0.04 \pm 0.01$ &  $0.16 \pm 0.01$ & $0.06 \pm 0.01$ &  $1.96 \pm 0.08$ \\
  {[\ion{Na}{vi}]}   &   14.3964 & 138.40 & $8.242 \times 10^4$ &  $0.12 \pm 0.02$ & $0.008 \pm 0.002$ &  $< 0.03$ &  $0.06 \pm 0.01$ & $< 0.02$ &  $0.11 \pm 0.03$ \\
  {[\ion{Ne}{iii}]}  &   15.5551 &  40.96 & $2.088 \times 10^5$ &  $5.42 \pm 0.05$ & $0.60 \pm 0.01$ &  $0.36 \pm 0.02$ &  $3.26 \pm 0.05$ & $0.27 \pm 0.01$ & $31.64 \pm 0.05$ \\
  {[\ion{Ne}{iii}]$_\text{b}$} & 15.5551 & 40.96 & $2.088 \times 10^5$ &        -- &              -- &               -- &  $0.12 \pm 0.02$ &              -- &               -- \\
  {[\ion{P}{iii}]}   &    17.885 &  19.77 & $3.915 \times 10^4$ &  $0.36 \pm 0.05$ &        $< 0.03$ &         $< 0.12$ & $0.16 \pm 0.07$ & $0.05 \pm 0.01$ & $1.77 \pm 0.05$ \\
  {[\ion{Fe}{ii}]}   &  17.93595 &   7.90 & $5.277 \times 10^4$ &  $0.29 \pm 0.05$ &        $< 0.03$ &         $< 0.15$ &  $0.11 \pm 0.06$ &        $< 0.09$ &  $1.64 \pm 0.06$ \\
  {[\ion{S}{iii}]}   &    18.713 &  23.34 & $1.188 \times 10^4$ &  $6.11 \pm 0.11$ & $0.43 \pm 0.02$ &  $1.02 \pm 0.09$ &  $3.94 \pm 0.20$ & $1.74 \pm 0.06$ & $45.76 \pm 0.11$ \\
  {[\ion{S}{iii}]$_\text{b}$} & 18.713 & 23.34 & $1.188 \times 10^4$ &          -- &              -- &               -- & $0.06 \pm 0.01$ &        -- &               -- \\
  {[\ion{Fe}{iii}]}  &    22.925 &  16.19 & $3.690 \times 10^4$ &  $0.25 \pm 0.13$ &        $< 0.09$ &         $< 0.12$ & $0.19 \pm 0.07$ &  $< 0.24$ &  $1.79 \pm 0.26$ \\
  {[\ion{Ne}{v}]}    &   24.3175 &  97.12 & $5.952 \times 10^3$ &  $2.79 \pm 0.24$ & $0.55 \pm 0.02$ &         $< 0.27$ & $2.60 \pm 0.26$ &         $< 0.21$ & $16.55 \pm 0.55$ \\
  {[\ion{Ne}{v}]$_\text{b}$} & 24.3175 & 97.12 & $5.952 \times 10^3$ &          -- &              -- &               -- & $0.16 \pm 0.01$ &        -- &               -- \\
  {[\ion{O}{iv}]}    &   25.8903 &  54.94 & $9.905 \times 10^3$ &  $6.30 \pm 0.35$ & $1.56 \pm 0.09$ & $0.33 \pm 0.07$ & $6.19 \pm 0.28$ &   $< 0.27$ & $45.90 \pm 0.95$ \\
  {[\ion{O}{iv}]$_\text{b}$} & 25.8903 & 54.94 & $9.905 \times 10^3$ &          -- &              -- &               -- & $0.34 \pm 0.14$ &               -- &               -- \\
  {[\ion{Fe}{ii}]}   &  25.98829 &   7.90 & $1.282 \times 10^4$ &  $0.52 \pm 0.15$ &        $< 0.12$ & $0.15 \pm 0.07$ & $0.29 \pm 0.14$ & $0.16 \pm 0.05$ & $4.86 \pm 0.59$ \\
  \hline
\end{tabular}
\tablefoot{Critical densities obtained with \textsc{PyNeb} \citep{luridiana15}, assuming LTE with $T_\text{e} = 10\,000\, \mathrm{K}$. For the fast ionised gas stream (\#3), fluxes corresponding to the broad blueshifted component $-1200 < v < -500\, \mathrm{\kms}$ are indicated by the subscript `b'.}
\end{table*}
\end{landscape}
\twocolumn


\onecolumn
\begin{figure*}[ht!!!]
\section[Rotational diagrams for warm H2 transitions]{Rotational diagrams for warm H$_2$ transitions}\label{app_H2}
  \centering
  \subfigure[]{\includegraphics[width = 0.49\columnwidth]{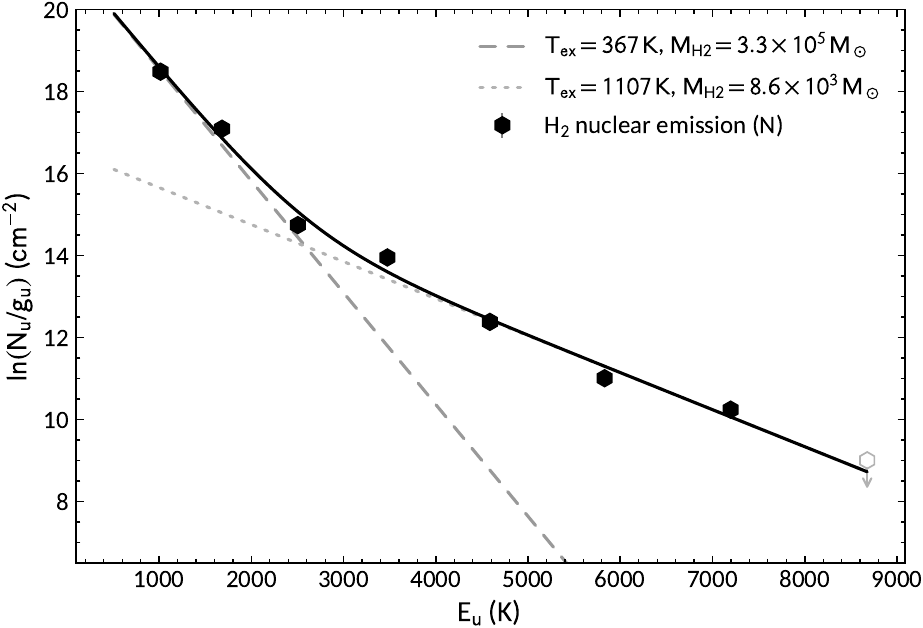}\label{subfig_H2_nucl}}~
  \subfigure[]{\includegraphics[width = 0.49\columnwidth]{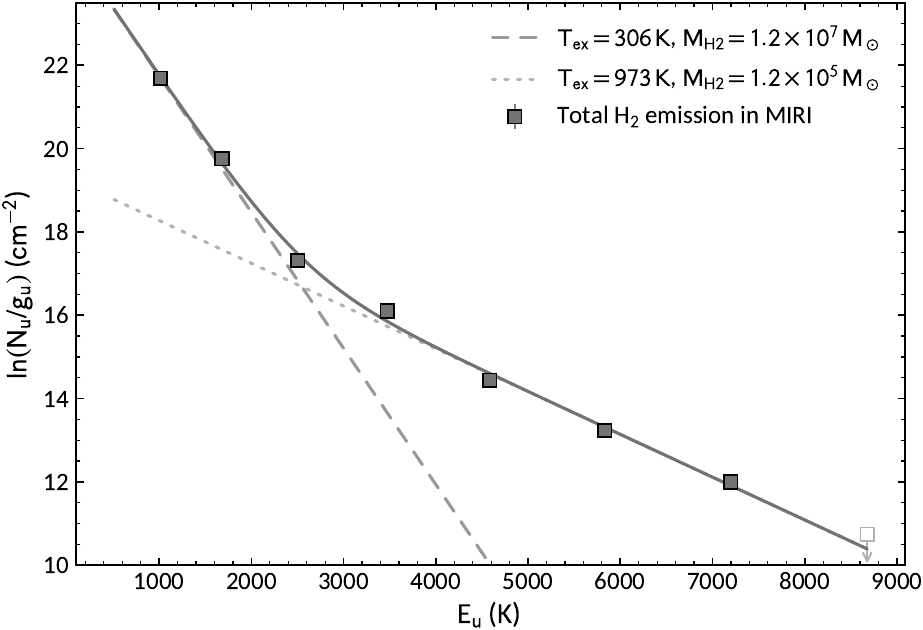}\label{subfig_H2_all}}\\[-0.2cm]
  \subfigure[]{\includegraphics[width = 0.49\columnwidth]{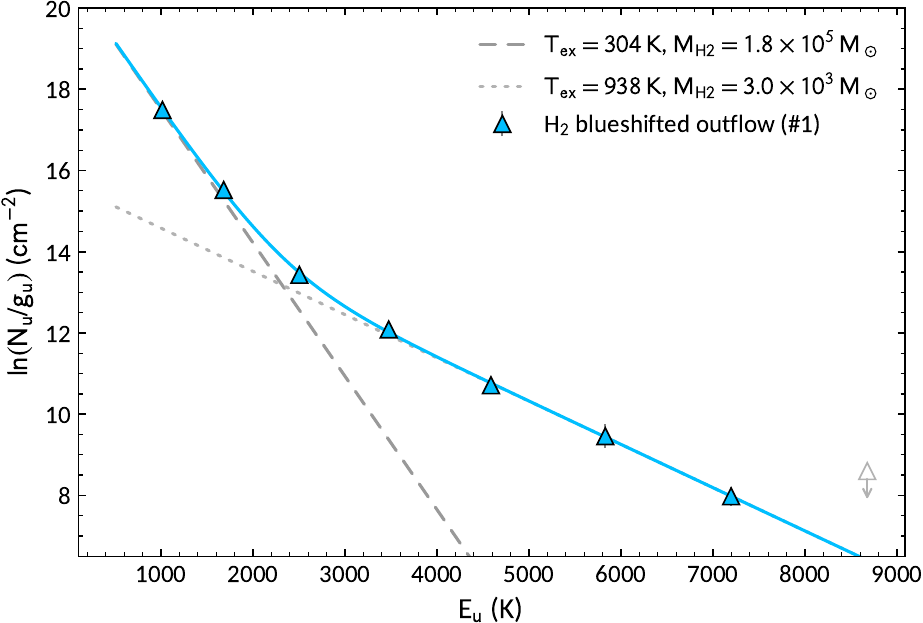}\label{subfig_H2_blue}}~
  \subfigure[]{\includegraphics[width = 0.49\columnwidth]{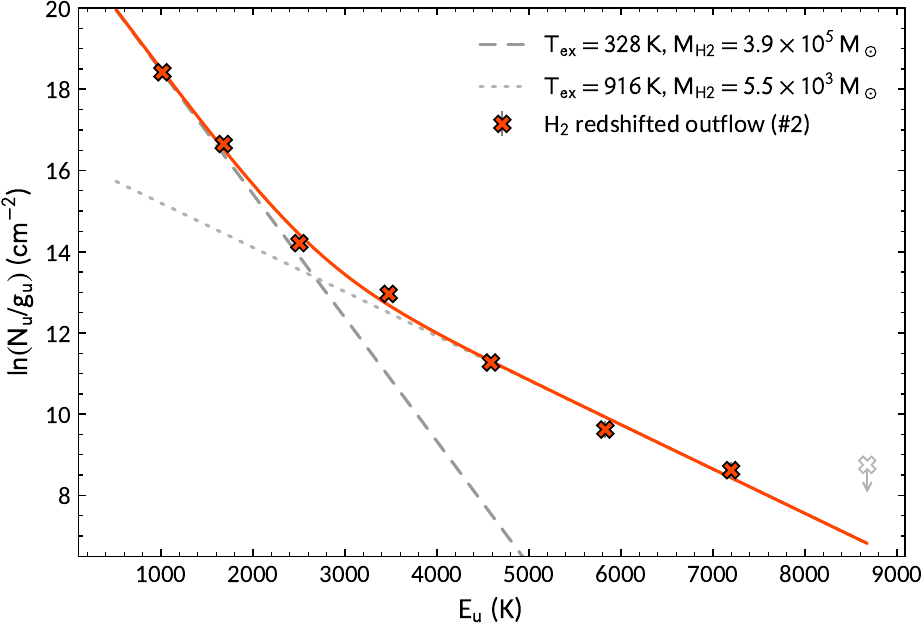}\label{subfig_H2_red}}\\[-0.2cm]
  \subfigure[]{\includegraphics[width = 0.49\columnwidth]{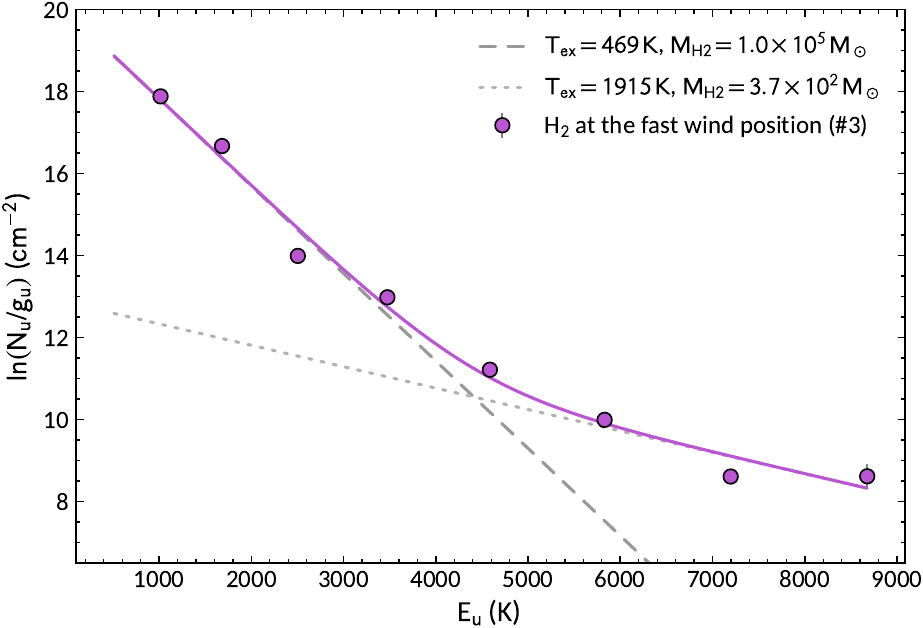}\label{subfig_H2_wind}}~
  \subfigure[]{\includegraphics[width = 0.49\columnwidth]{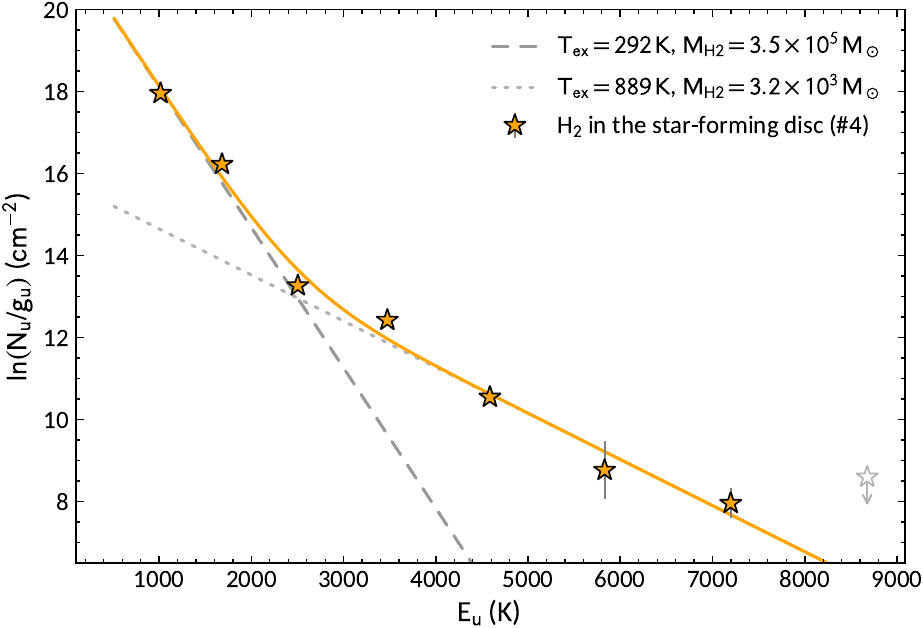}\label{subfig_H2_disc}}\\[-0.2cm]
  \caption{Rotational diagrams (Boltzmann plots) for the warm H$_2$ rotational transitions S(1) to S(8) detected in ESO\,420-G13, for the extracted nuclear spectra \textbf{(a)}, the integrated MIRI FoV \textbf{(b)}, and the rest of the apertures indicated in Fig.\,\ref{fig_regions} \textbf{(c--f)}. The horizontal axis shows the upper state energy for each transition ($E_\text{u}$), while the vertical axis shows the normalised upper state column density ($N_\text{u} / g_\text{u}$). We adopt an ortho-to-para ratio of 3, corresponding to LTE conditions at $T \gtrsim 200\,\mathrm{K}$ \citep{burton92,sheffer11}. In each panel, the solid line shows a fit of the normalised column density including a colder but more massive component (dashed line) and a hotter but less massive one (dotted line). Note that the highest warm H$_2$ temperatures are obtained for the aperture centred on the fast ionised gas stream, while the coldest temperatures are measured for the aperture centred on the star-forming region.}\label{fig_H2}
\end{figure*}

\FloatBarrier
\twocolumn


\onecolumn
\section{Ionised gas mass estimate from collisional excitation transitions}\label{app_ionmass}

The emissivity of a collisionally excited line is set by the rate of radiative decay between the two ionic levels involved in the transition. The level populations are determined by statistical equilibrium, which balances the collisional and radiative processes that populate and depopulate a given level $i$ \citep{osterbrock06,luridiana15}:
\begin{equation} \label{eq_colleq}
    \begin{split}
     &\aoverbrace[L1R]{\sum_{j \ne i} n_\text{e} n_j q_{ji}}[L1A1R]^{\substack{\text{col. excitation} \\ \text{to $i$-level}}} + \aoverbrace[L1R]{\sum_{j>i} n_j A_{ji}}[L1A1R]^{\substack{\text{rad. decays}\\ \text{to $i$-level}}} = \aoverbrace[L1R]{\sum_{j \ne i} n_\text{e} n_i q_{ij}}[L1A1R]^{\substack{\text{col. deexcitation} \\ \text{from $i$-level}}} + \aoverbrace[L1R]{\sum_{j<i} n_i A_{ij}}[L1A1R]^{\substack{\text{rad. decays} \\ \text{from $i$-level}}} \\
     &\sum_{i} n_i = n_\text{ion} = f_\text{ion} \, n_\text{X} \ ; \quad \sum_\text{ion} n_\text{ion} = n_\text{X} = \left( \frac{X}{H} \right) \, n_\text{H}
\end{split}
\end{equation}

In the first line, level $i$ is populated by collisional transitions from other levels and radiative cascades from higher-lying levels. On the left side of the equation, the collisional term depends on the electron density ($n_\text{e}$), the density of ions in the initial energy level ($n_j$), and the collisional excitation rate between the two levels ($q_{ji}$). The radiative term depends on the density of ions in the upper level ($n_j$) and the Einstein coefficient for spontaneous emission to the lower level ($A_{ji}$). On the right side, level $i$ is depopulated by collisional deexcitations to other levels ($n_\text{e} n_i q_{ij}$) and spontaneous decays to lower levels ($n_i A_{ij}$). The second line of the equation represents particle conservation, i.e. the sum of level populations equals the total ion density ($n_\text{ion}$), which can be written in terms of the ionic fraction ($f_\text{ion}$) and the total elemental density ($n_\text{X}$). The latter is expressed via the elemental abundance $(X/H)$ relative to the hydrogen density ($n_\text{H}$). On the other hand, the collisional deexcitation ($q_{ij}$) and excitation rates ($q_{ji}$) can be derived from the velocity-averaged collision strengths $\Upupsilon_{ij}$, which depend on the electron temperature \citep{osterbrock06}:
\begin{equation}\label{eq_qij}
  \begin{split}
  q_{ij} &= \frac{h^2}{(2 \pi \, m_\text{e})^{3/2}} \frac{\Upupsilon_{ij}}{g_i \sqrt{k T_\text{e}}} = \frac{8.629 \times 10^{-6}}{g_i \sqrt{T_\text{e}}} \, \Upupsilon_{ij} \\
  q_{ji} &= \frac{g_i}{g_j} \, q_{ij} \, e^{-h \nu_{ij} / k T_\text{e}} = \frac{8.629 \times 10^{-6}}{g_j \sqrt{T_\text{e}}} \, \Upupsilon_{ij} \, e^{-h \nu_{ij} / k T_\text{e}}
  \end{split}
  \hfill (i > j) \quad
\end{equation}
where $g_j$ and $g_i$ are the statistical weights of levels $j$ and $i$, respectively, and $h \nu_{ij}$ is the energy difference between the levels (i.e. the transition energy). $T_\text{e}$ is the electron temperature, $k$ is the Boltzmann constant, $h$ is the Planck constant, and $m_\text{e}$ is the electron mass.

\begin{figure}[ht!]
\centering
\includegraphics[width = 0.55\textwidth]{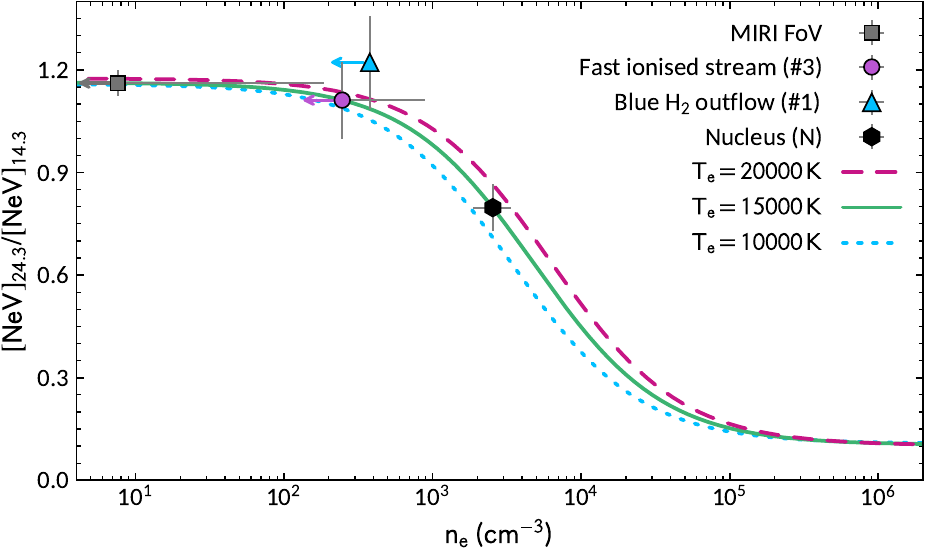}
\caption{Electron densities estimated for the [\ion{Ne}{v}]$_{24.3}$/[\ion{Ne}{v}]$_{14.3}$ ratios measured in different apertures (see Fig.\,\ref{fig_regions}) and the theoretical relation from \textsc{PyNeb} \citep{luridiana15}. The estimated densities are interpolated for the curve with an electron temperature of $15\,000\, \mathrm{K}$ (solid green line), while the relations for $20\,000\, \mathrm{K}$ (dashed purple line) and $10\,000\, \mathrm{K}$ (dotted blue line) are also shown for comparison.}\label{fig_dens}
\end{figure}

To derive the rate of radiative decays for a certain transition, one must solve the set of equations in Eq.\,\ref{eq_colleq} for all levels that potentially contribute to the population of level $i$. In practice, the calculations may only take into account the lower levels, i.e. those close to the ground state that can be significantly populated by collisions. Additionally, the equations are simplified when the gas density is below the critical densities of the transitions used. In this case, collisions are rare and do not contribute significantly to the gas deexcitation, which is then dominated by radiative decays. In this context, the critical densities for all the neon transitions in Table\,\ref{tab_ionflux} are above $6 \times 10^3\, \mathrm{cm^{-3}}$ at $\gtrsim 10^4\, \mathrm{K}$, while the estimated densities from the [\ion{Ne}{v}]$_{24.3/14.3}$ ratio are below the low-density limit ($n_\text{e} \lesssim 300\, \mathrm{cm^{-3}}$), except for the nuclear aperture where higher densities are measured ($\sim 2500\, \mathrm{cm^{-3}}$; Fig.\,\ref{fig_dens}).

\begin{figure*}[ht!]
\centering
\subfigure[]{\includegraphics[width = 0.5\columnwidth]{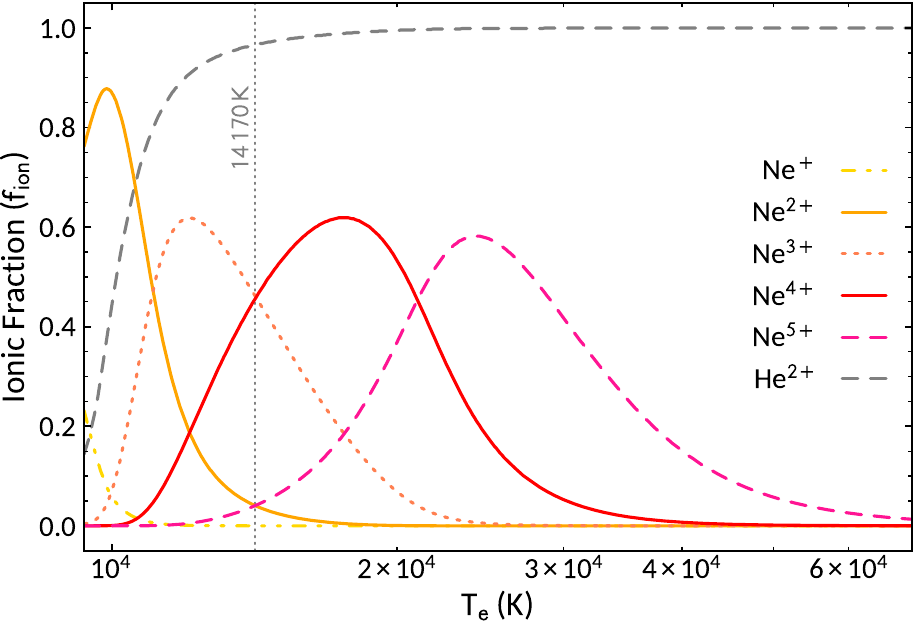}\label{subfig_ionfrac}}~
\subfigure[]{\includegraphics[width = 0.5\columnwidth]{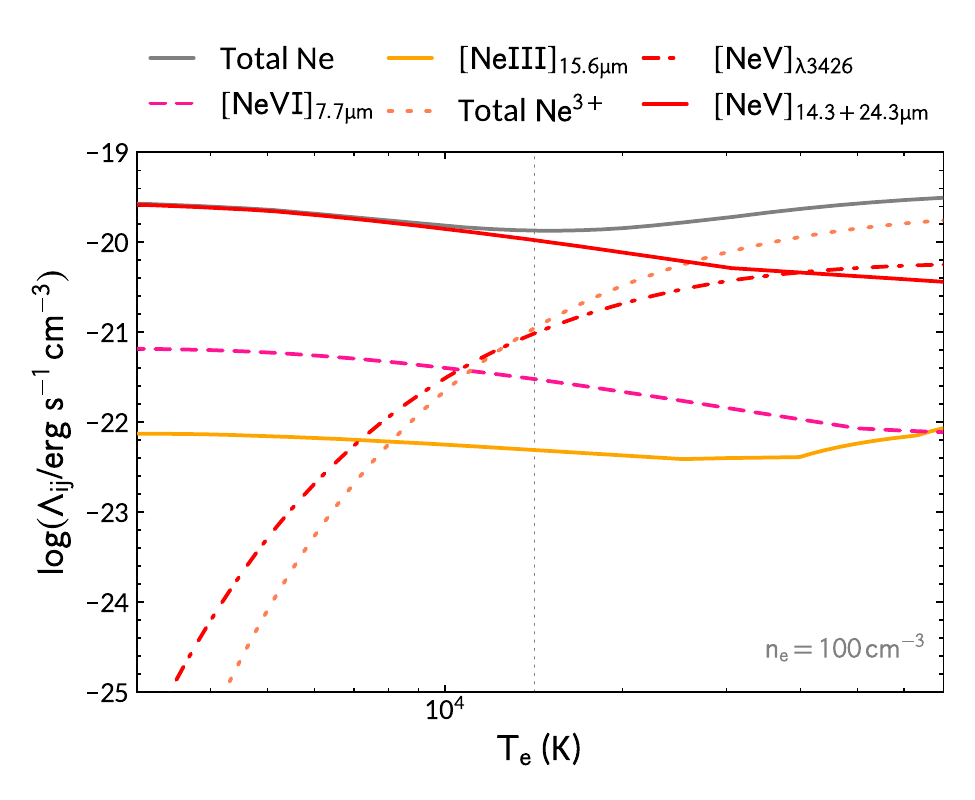}\label{subfig_emiss}}\\[-0.2cm]
\caption{\textbf{(a)} expected ionic fractions of the different neon species (coloured lines) and He$^{2+}$ (dashed grey line) as a function of the electron temperature for AGN photoionisation using \textsc{Cloudy} models (see Section\,\ref{disc_coronal}; \citealt{chatzikos23}). The most abundant neon species for a $\sim 15\,000\, \mathrm{K}$ plasma are Ne$^{3+}$ and Ne$^{4+}$. \textbf{(b)} line emissivity for the main optical/UV and mid-IR transitions as a function of the electron temperature derived with \textsc{PyNeb} \citep{luridiana15}, adopting the ionic fractions predicted in \textbf{(a)} for a temperature of $15\,000\, \mathrm{K}$.}\label{fig_emiss}
\end{figure*}

\pagebreak
The cooling rate per unit volume for a transition from upper level $i$ to lower level $j$, $\Lambda_{ij}$, can be derived from the radiative decay rate and the transition energy by considering the population balance of level $i$. This includes collisional transitions from any other level into level $i$ ($n_{\rm e}n_k q_{ki}$) and radiative cascades from higher levels ($n_k A_{ki}$, for $k>i$), as well as collisional de-excitations from level $i$ to any other level ($n_{\rm e}n_i q_{ik}$) and radiative decays from level $i$ to lower levels other than $j$ ($n_i A_{ik}$, for $k<i$ and $k\ne j$):
\begin{equation}\label{eq_cool}
  \Lambda_{ij} = n_i A_{ij} h \nu_{ij} = h \nu_{ij} \left( \sum_{k \ne i} n_\text{e} n_k q_{ki} \ + \sum_{k > i} n_k A_{ki} \ - \underbrace{\sum_{k \ne i} n_\text{e} n_i q_{ik}}_{\simeq \, 0 \ \text{for} \ n_\text{e} < n_\text{crit}} \ - \ \sum_{k < i, k \ne j} n_i A_{ik} \right ) \hfill (i > j) \quad
\end{equation}

\textsc{PyNeb} \citep{luridiana15} solves Eq.\,\ref{eq_qij} for an $n$-level atom and provides normalised line emissivities $\gamma_{ij}$ for given temperature and density. Once the cooling rates per unit volume are known, the corresponding luminosity $L_{ij}$ from a given volume $V$ with filling factor $\epsilon$ can be derived. The luminosity is then related to the observed line flux $F_{ij}$ through the luminosity distance $D_\text{L}$:
\begin{equation}\label{eq_fij}
  \begin{split}
  \gamma_{ij} &= \frac{n_i}{n_\text{ion} n_\text{e}} A_{ij} h \nu_{ij} \\
  L_{ij} &= \Lambda_{ij} \, V \epsilon = n_\text{ion} n_\text{e} \, \gamma_{ij} \, V \epsilon = f_\text{ion} \frac{X}{H} n_\text{H} \, n_\text{e} \, \gamma_{ij} \, V \epsilon = 4 \pi D_\text{L}^2 \, F_{ij}
  \end{split}
  \hfill (i > j) \quad
\end{equation}

On the other hand, the total ionised gas mass within the nebula $M_\text{\sc Hii}$ can be approximated by the combined hydrogen and helium mass, which depends on the hydrogen and helium densities and the effective volume $V \epsilon$:
\begin{equation}\label{eq_mass1}
  M_\text{\sc Hii} = (n_\text{H} m_\text{p} + n_\text{He} m_\text{He}) \, V \epsilon = m_\text{p} (n_\text{H} + 4 n_\text{He}) \, V \epsilon = m_\text{p} n_\text{H} \left(1 + 4 \frac{\text{He}}{H} \right) V \epsilon \sim 1.33 \, m_\text{p} n_\text{H} \, V \epsilon
\end{equation}
where $m_\text{p}$ is the proton mass, $m_\text{He} \sim 4 m_\text{p}$ is the helium mass, and the helium density $n_\text{He}$ can be approximated as $\sim 0.082 \, n_\text{H}$, assuming a solar (He/H) abundance ratio \citep{asplund21}. Thus, from Eqs.\,\ref{eq_fij}, and \ref{eq_mass1}, the total ionised gas mass can be expressed as:
\begin{equation}\label{eq_mass2}
  M_\text{\sc Hii} = \frac{1.33 \, m_\text{p}}{f_\text{ion} (X/H) n_\text{e}} \, \frac{4 \pi D_\text{L}^2 \, F_{ij}}{\gamma_{ij}} \hfill (i > j) \quad
\end{equation}
Note that this equation no longer depends on the gas volume or filling factor, though the line emissivity retains an implicit dependence on electron temperature. Thus, Eq.\,\ref{eq_mass2} provides an estimate of the total ionised gas mass as a function of the observed flux for a collisionally excited transition, the electron temperature and density, the elemental abundance and the ion relative abundance. Nevertheless, robust mass estimates require that the observed transition $F_{ij}$ be produced by a dominant ionic species that contributes significantly to cooling. Otherwise, uncertainties in the ionic fraction and electron temperature propagate into $M_\text{\sc Hii}$. To minimise these uncertainties, multiple transitions can be used to compute the ionised gas mass. In this case, Eq.\,\ref{eq_mass2} can be generalised as:
\begin{equation}\label{eq_mass3}
  M_\text{\sc Hii} = \frac{1.33 \, m_\text{p}}{(X/H) n_\text{e}} \, 4 \pi D_\text{L}^2 \, \frac{\sum_{ij} F_{ij}}{\sum_{ij} f^{ij}_\text{ion} \, \gamma_{ij}} \hfill (i > j) \quad
\end{equation}
where $f_\text{ion}$ may vary if transitions from different ions are considered. Figure\,\ref{fig_emiss}a shows the expected ionic fractions of different neon species as a function of temperature for an AGN photoionised nebula, simulated with \textsc{Cloudy} v23.01 \citep{chatzikos23}. We used the default AGN template from \citet{mathews87} as the input continuum, assuming constant density of $300\,\mathrm{cm^{-3}}$ and plane-parallel geometry. The detection of collimated ionised wind in both [\ion{Ne}{iii}]$_{15.6}$ and [\ion{Ne}{vi}]$_{7.7}$ lines (Fig.\,\ref{fig_windspec}) is consistent with a relatively narrow temperature range, as shown in Fig.\,\ref{fig_emiss}a. Thus, we adopt $T_\text{e} \sim 15\,000\,\mathrm{K}$ to estimate the neon ionic fractions. Figure\,\ref{fig_emiss}b shows the predicted emissivities for the main optical and IR neon transitions, obtained with \textsc{PyNeb} \citep{luridiana15} for a density of $300\,\mathrm{cm^{-3}}$, adopting the previously derived ionic fractions. These estimates suggest that for an AGN-ionised nebula at $T_\text{e} \lesssim 20\,000\,\mathrm{K}$, the two mid-IR Ne$^{4+}$ lines are by far the main coolants among all neon transitions. This is due to the large Ne$^{4+}$ ionic fraction expected under such conditions ($\sim 46\%$) and the lack of energy levels close to the ground state --\,and therefore of mid-IR transitions\,-- for Ne$^{3+}$, which also accounts for almost half of the ionic fraction. The cooling contribution from optical transitions of Ne$^{3+}$ and Ne$^{4+}$ only becomes important at very high temperatures ($\gtrsim 20\,000\,\mathrm{K}$), when the corresponding energy levels become sufficiently populated.

Finally, the form of Eq.\,\ref{eq_mass3} when the two [\ion{Ne}{v}]$_{14.3,24.3}$ transitions are taken into account can be expressed as a function of the neon abundance (Ne/H) and the Ne$^{4+}$ ionic fraction ($f_\mathrm{Ne^{4+}}$):
\begin{equation}\label{eq_mass_ne_abund}
  M_\text{\sc Hii} = 4.3 \times 10^{12} f_\mathrm{Ne^{4+}} \left( \frac{\text{Ne}}{H} \right) \left( \frac{n_\text{e}}{\mathrm{cm^{-3}}} \right)^{-1} \left( \frac{D_\text{L}}{\mathrm{Mpc}} \right)^2 \left( \frac{F_{14.3+24.3}}{\mathrm{erg\,cm^{-2}\,s^{-1}}} \right) \, \mathrm{M_\odot}
\end{equation}


\section{Ionised and molecular gas outflow properties}\label{app_outflow_props}

This appendix summarises the mass and energy budgets derived for the molecular and ionised outflows in ESO\,420-G13. The quantities listed in Table\,\ref{tab_winds} correspond to the two launching scenarios discussed in Section\,\ref{disc_bubble}, namely whether the molecular outflows originate at the position of the fast ionised gas stream (region~\#3) or directly from the nucleus. For each case, we provide the adopted outflow radius and velocity, gas mass, mass outflow rate, momentum rate, and kinetic luminosity, allowing a direct comparison between the different gas phases and geometrical assumptions.

\begin{table*}[h!]
\caption{Energy and momentum budget of the ionised and molecular gas outflows.}\label{tab_winds}
\centering
\setlength{\tabcolsep}{5.0pt}
\begin{tabular}{lccccccc}
  & \bf $r_\text{out}$ & \bf $v_\text{out}$ & \bf $\sigma_\text{out}$ & \bf M$_\text{out}$ & \bf $\dot{M}_\text{out}$ & \bf $\dot{M}_\text{out} v_\text{out}$ & \bf L$_\text{kin} = \frac{1}{2} \dot{M}_\text{out} (v^2_\text{out} + 2\,\sigma^2_\text{out})$ \\
             & [pc]  & [$\mathrm{\kms}$] & [$\mathrm{\kms}$] & [$\mathrm{M_\odot}$] & [$\mathrm{M_\odot\,yr^{-1}}$] & [$\mathrm{erg\,cm^{-1}}$] & [$\mathrm{erg\,s^{-1}}$] \\
  \hline \\[-0.1cm]
  \bf Region & \multicolumn{7}{c}{Scenario 1: outflows launched from region \#3 in the northern jet} \\
  \hline \\[-0.3cm]
  Blueshifted H$_2$ outflow (\#1) & 400  & -70 & 60 & $4.0\times 10^6$ & $2.2$ & $1.0 \times 10^{33}$ & $0.8 \times 10^{40}$ \\
  Redshifted H$_2$ outflow (\#2)  & 380  & 160 & 70 & $8.7\times 10^6$ & $11.2$ & $11.3 \times 10^{33}$ & $12.5 \times 10^{40}$ \\
  Fast ionised gas stream (\#3)   & 210 & -750 & 210 & $5.1\times 10^3$ & $0.06$ & $2.6 \times 10^{32}$ & $1.1 \times 10^{40}$ \\
  \hline \\[-0.1cm]
  \bf Region & \multicolumn{7}{c}{Scenario 2: outflows launched from the nucleus} \\
  \hline\\[-0.3cm]
  Blueshifted H$_2$ outflow (\#1) & 660  & -70 & 60 & $4.0\times 10^6$ & $1.3$ & $0.6 \times 10^{33}$ & $0.5 \times 10^{40}$ \\
  Redshifted H$_2$ outflow (\#2)  & 450  & 160 & 70 & $8.7\times 10^6$ & $9.5$ & $9.6 \times 10^{33}$ & $10.6 \times 10^{40}$ \\
  Collimated coronal gas          & 870 & -80 & 80 & $4.8 \times 10^5$ & $0.14$ & $6.9 \times 10^{31}$ & $8.2 \times 10^{38}$ \\
  \hline
\end{tabular}
\tablefoot{Region name, distance to launching point assuming region \#3 or the nucleus as the origin of the outflow, outflow velocity ($v_\text{out}$) and velocity dispersion ($\sigma_\text{out}$), total outflow gas mass M$_\text{out}$ (M$^\text{warm}_\text{H2}$ + M$^\text{cold}_\text{H2}$ for regions \#1 and \#2; M$_\text{\sc Hii}$ for the broad high-ionisation component in region \#3 and the total extended ionised outflow), mass outflow rate ($\dot{M}$), momentum rate ($\dot{M}v$), kinetic luminosity ($L_\text{kin}$).}
\end{table*}

\twocolumn
\end{appendix}
\end{document}